\documentclass[journal]{IEEEtran}
\usepackage{graphicx}
\usepackage{color}
\usepackage{fancybox}
\usepackage{tcolorbox}
\usepackage{cite}
\usepackage{amsmath,amssymb}
\usepackage{siunitx}
\usepackage{comment}
\usepackage{algorithm}
\usepackage[noend]{algpseudocode}
\usepackage{multirow}
\usepackage{dblfloatfix}
\usepackage{xcolor}
\usepackage{caption}
\usepackage{amsthm}
\usepackage{bbold}
\usepackage{bm}
\newtheorem{theorem}{Theorem}
\newtheorem{corollary}{Corollary}
\newtheorem{assumption}{Assumption}
\newtheorem{remark}{Remark}
\newtheorem{lemma}{Lemma}
\newtheorem{definition}{Definition}
\newtheorem{problem}{Problem}

\begin{document}
\renewcommand{\arraystretch}{1.15}

\title{{Network Topology Invariant} Stability Certificates for DC Microgrids with Arbitrary Load Dynamics}

\author{Samuel Chevalier,~\IEEEmembership{Student Member,~IEEE,}
        {Federico Martin Ibanez,~\IEEEmembership{Senior Member,~IEEE,}}
        Kathleen Cavanagh,~\IEEEmembership{Student Member,~IEEE,}
        Konstantin Turitsyn,~\IEEEmembership{Member,~IEEE,}
        Luca Daniel,~\IEEEmembership{Member,~IEEE,} and\\
        Petr Vorobev,~\IEEEmembership{Member,~IEEE,}
        
\thanks{{This work was supported in part by the Skoltech-MIT Next Generation grant and by the MIT Energy Initiative Seed Fund Program.}

{S. Chevalier and K. Cavanagh were with the Department of Mechanical Engineering, Massachusetts Institute of Technology, Massachusetts, USA. E-mail: schev@mit.edu, kcav@mit.edu.}

{F. Ibanez is with the Skolkovo Institute of Science and Technology (Skoltech), Moscow, Russia. E-mail: fm.ibanez@skoltech.ru.}

{K. Turitsyn is with the D. E. Shaw Group, New York, New York, USA. E-mail: turitsyn@mit.edu.}

{L. Daniel is with the Department of Electrical Engineering and Computer Science, Massachusetts Institute of Technology, Massachusetts, USA. E-mail: luca@mit.edu.}

{P. Vorobev is with the Skolkovo Institute of Science and Technology (Skoltech), Moscow, Russia. E-mail: p.vorobev@skoltech.ru.}}}

\renewcommand{\baselinestretch}{.95}

\maketitle


\begin{abstract}
DC microgrids are prone to small-signal instabilities due to the presence of tightly regulated loads. This paper develops a decentralized stability certificate which is capable of certifying the small-signal stability of an islanded DC network containing such loads. Utilizing a novel homotopy approach, the proposed standards ensure that no system eigenmodes are able to cross into the unstable right half plane for a continuous range of controller gain levels. The resulting ``standards" can be applied to variety of grid components which meet the specified, but non-unique, criteria. These standards thus take a step towards offering \emph{plug-and-play} operability of DC microgrids. The proposed theorems are explicitly illustrated and numerically validated on {multiple DC microgrid test-cases containing both buck and boost converter dynamics}.
\end{abstract}


\IEEEpeerreviewmaketitle


\section{Introduction}\label{sec:introduction}
\IEEEPARstart{R}{ecent} advances in power electronics technologies and the general trend towards renewable energy sources have lead to increased interest in DC grids \cite{planas2015ac,justo2013ac}. Small-scale DC microgirds have been in use for several decades already, mainly as autonomous electric systems on board of vehicles \cite{emadi2003vehicular}. As such, the configuration of these microgrids was fixed and well planned for the exact operating conditions which were known in advance. On the other hand, DC microgrids with an ``open" structure, capable of being expanded and reconfigured while remaining stable for a broad range of loading conditions, have mostly been out of the scope of academic research. In part, this is justified by the fact that, currently, the majority of microgrids are using an AC interface for power distribution, even if all the sources and loads are naturally DC. However, fully DC microgrids can become an economically feasible solution for supplying power in remote communities.

Unlike in AC grids where a substantial part of the load is of the electro-mechanical type, loads in DC grids are mostly represented by power electronics converters with tight controls to achieve flat voltage at their outputs \cite{erickson2007fundamentals}. This leads to a constant power load (CPL) behavior on the input within the control loop bandwidth, which is regarded to be one of the main sources of instabilities in DC grids. The origin of the instability is often related to the, so-called, negative incremental resistance introduced by CPL; a number of methods for stability assessment are based on such representation. Recent reviews \cite{singh2017constant,riccobono2012comprehensive} present a comprehensive classification of the existing stability criteria and stabilization methods for DC grids. 

Recently, the problem of stability assessment for CPL-based microgrids has attracted substantial attention from the controls community~\cite{Sanchez:2013gl, SimpsonPorco:2015hp, Zhao:2015eu, Barabanov:2016ki, de2018power}.
In our previous works on the subject \cite{belk2016stability, inam2016stability, Cavanagh:2017uo}, we have demonstrated that the problem of linear and transient stability of DC microgrids with CPL can be addressed using Brayton-Moser mixed potential approach \cite{Brayton:1964gr, Jeltsema:2009jd}. However, stability conditions derived under the conservative CPL modeling assumption may lead to excessive constraints on network configuration and installed equipment, and they neglect the complex dynamical behaviour of the controllers which regulate these loads. In reality, regulated power converters act similar to CPL only within their control loop bandwidth~\cite{rahimi2009analytical}. Accordingly, methods which explicitly account for the finite bandwidth of load controllers can potentially provide less restrictive stability conditions. However, dealing with load models that are more complex than a simple CPL requires development of rather specific techniques. 

Traditionally, the power electronics community has relied on a number of different impedance based stability conditions. {An excellent overview and historical perspective on such impedance based methods can be found in the recent review article~\cite{Shah:2021}}. Most of these methods consider the minor loop gain, which is the ratio of the source output impedance to the load input impedance. The celebrated Middlebrook criterion \cite{MB1976}, originally proposed for input filter design, is based on a small-gain condition for minor loop gain, demanding its absolute value to be less than unity for all frequencies. While a rather conservative method, it only requires knowledge of the absolute values of impedances over the whole frequency range. A less conservative Gain-Margin Phase-Margin criterion \cite{wildrick1995method} allows the Nyquist plot of the minor loop gain to leave the unit circle provided there is a sufficient phase margin. Another method - the opposing argument criterion - is based on conditions imposed on the real part of the minor loop gain \cite{feng2002impedance}. The main advantage of this method is that it can be applied to multi-load systems, since the contributions from each individual load can now be simply added together. Finally, the least conservative methods are based on the so-called energy source consortium analysis (ESCA) \cite{sudhoff2000admittance} and the similar root exponential stability criterion (RESC) \cite{sudhoff2011advancements}. Both offer the smallest forbidden region for the minor loop gain of all the existing methods.

{There exist a number of stability assessment approaches in the literature which require more centralized coordination and analysis. The authors in~\cite{Fan:2020} compute the eigenmodes of the centralized network admittance function and show why eigen-based approaches can be advantageous over Nyquist and bode-based techniques. In~\cite{Gu:2015}, the authors propose the use of a ``self-disciplined stabilization" approach, whereby all controllers are intrusively altered to meet a particular passivity condition. In cases were converter parameters are uncertain but known to live in some definite range,~\cite{Lucas:2019} utilizes Kharitonov’s theorem inside of a linear programming formulation in order to develop robust control laws. Similarly,~\cite{Liu:2018} defines a polytopic uncertainty set associated with network parameter and equilibrium point uncertainties; they then pose the associated ``robust stability" problem as an equivalent convex optimization problem. In order to overcome the negative incremental resistance provided by CPLs, a number of works~\cite{Zhang:2015,Hamzeh:2016,Kim:2016} have proposed the use of so-called ``virtual impedance" as a counteracting control tool. The tuning of this virtual impedance can take many forms. Most recently,~\cite{Adly:2021} proposed the use of virtual resistance to ensure stability. However, the approach which is used for selecting the size of this virtual resistance assumes that the loads exhibit CPL behaviour across all frequencies; this assumption can lead to overly conservative control laws. Furthermore, this method assumes prior knowledge of all devices which will be interconnected near the loads.}

Most of the described stability assessment methods are { either (i) centralized in nature, or (ii) primarily concerned with single-source single-load systems and a possible destabilizing interaction between the impedances of corresponding components.} Under certain conditions, it could be possible to generalize some these {latter} methods by attributing network components to either loads or sources and changing the corresponding effective impedance. In all the cases, though, explicit knowledge of both the models of all the system components and the network configuration are required to perform the stability assessment. However, in the case of the aforementioned ``open structure" DC grids, not all of the parameters can be exactly known. Moreover, running a separate {centralized} stability test upon every grid reconfiguration can be practically infeasible. Thus, there is the need for stability assessment methods that can be realized under limited knowledge about the system parameters and its configuration, {and that simultaneously are not overly conservative or intrusive.}

In the present manuscript, {we propose a stability assessment tool which fulfills these requirements and is therefore unique in the academic literature on DC microgrid stability assessment.} Accordingly, we have made a step towards offering a ``plug-and-play" approach for DC grids, where certain general criteria can be issued for both loads and network elements that can guarantee stable operation of the full system under arbitrary grid configurations. {Using decentralized positive realness conditions}, we develop a specialized procedure that allows for the overall system stability assessment to be reduced to a separate consideration for individual load and line impedances, thus resulting in completely decentralized stability criteria.




The main contributions of the manuscript are as follows:

\begin{enumerate}
    \item {We develop a fully \emph{network topology invariant}} stability criteria which can be applied to DC grids with different components (i.e., sources, lines, and loads) and can certify stability for an arbitrary network configuration with these components. 
    \item We prove that any component which satisfies this criteria can be safely added to a network without a system eigenmode crossing into the unstable right half plane, thus allowing for “plug-and-play” operability of DC microgrids.
\end{enumerate}

{The remainder of the paper is structured as follows. In Section \ref{sec:formulation}, we review the relevant technical background by first stating a useful Laplace domain microgrid model and then by presenting a common constant power load model (which will be used in later simulation). In Section \ref{sec:genap}, we derive the decentralized stability certificates and present an analytically tractable example of their implementation. In Section \ref{sec:num}, we offer numerical test results on simulated microgrid networks for multiple converter types. Next, in Section \ref{sec:control_adjust}, we demonstrate how the derived stability standards can be used to guide controller tuning. {In Section \ref{sec: RTS}, we validate the stability conclusions of the proposed approach using real-time simulations in the OPAL-RT environment.} Then, in Section \ref{sec: design_operation}, we review how the decentralized standards can be used in practical microgrid design and operational settings. Finally, we discuss the benefits and shortcomings of our proposed methodology in Section \ref{sec:discus}, and we offer concluding remarks in Section \ref{sec:con}.}


\section{Technical Background}\label{sec:formulation}
In this section, {the general problem of small-signal stability certification is motivated,} and then an appropriate Laplace domain DC microgird network model is stated. Finally, the small-signal model of a constant power DC load is recalled.

\subsection{Network Modeling for Small-Signal Stability Analysis}
{According to the IEEE-PES task force on microgrids~\cite{Farrokhabadi:2020}, stability comes in many flavors, and it can be assessed across varying time scales. Most stability definitions are classified along the lines of either small disturbance (i.e., small-signal) or large disturbance (i.e., transient) stability. Large disturbance stability is essential for ensuring the system will converge back to a stable equilibrium in finite time after a significant contingency; small-signal stability, however, ensures that the operating point itself is robust to small perturbations. Thus, small-signal stability is a necessary prerequisite for large disturbance stability, and it is the exclusive type of stability we consider in this paper. Our primary goal, though, is to develop small-signal stability criteria which, if satisfied by a set of devices, will be applicable across a wide range of microgrid topologies and configurations. Thus, we say the resulting criteria will be \textit{topology invariant}.}

\begin{figure}
\centering
\includegraphics[width=0.95\columnwidth]{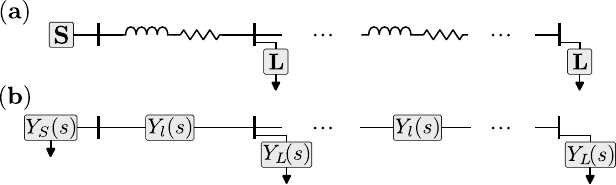}
\caption{\label{fig: MG_Power} {Panel (\textbf{a}) shows the general power scheme of a physical microgrid network, where loads ({\textbf L}) are connected to a source ({\textbf S}) (or potentially multiple sources) through electromagnetic lines. Panel (\textbf{b}) shows the Laplace domain equivalent circuit, where load dynamics $Y_L(s)$ interact with each other, and with the source dynamics $Y_S(s)$, through lines $Y_l(s)$.}}
\end{figure}

To study the small-signal stability of a DC microgrid, we leverage a network whose connected graph $G({\mathcal V},{\mathcal E})$ has edge set $\mathcal{E}$ with cardinality $|\mathcal{E}|=m$, vertex set $\mathcal{V}$ with cardinality $|\mathcal{V}|=n$, and signed nodal incidence matrix $E\in{\mathbb R}^{m\times n}$. This network has a set of operating equilibrium voltages ${\bf V}_0\in{\mathbb R}^n$ and nodal current injections ${\bf I}_0\in{\mathbb R}^n$. {An example of such a system is given by panel (\textbf{a}) of Fig. \ref{fig: MG_Power}, where a number of loads (denoted as \textbf{L}'s) are connected to a source (\textbf{S}) by a feeder line. In order to study the small-signal dynamics of the system, we introduce} the perturbations of the vectors of voltages and currents as ${\bf v}(t)={\bf V}(t)-{\bf V}_0$ and ${\bf i}(t)={\bf I}(t)-{\bf I}_0$, respectively. Here ${\bf V}_0$ and ${\bf I}_0$ are the steady-state values of the voltage and current vectors. Finally, the Laplace domain representations of the voltage and current perturbations are written as ${\boldsymbol v}(s) =  {\mathcal L} \{{\bf v}(t)\}$ and ${\boldsymbol i}(s) =  {\mathcal L} \{{\bf i}(t)\}$. {Fig. \ref{fig: MG_Power} (\textbf{b}) shows the Laplace domain representation of the microgrid given in Fig. \ref{fig: MG_Power} (\textbf{a}).}

Next, we define the frequency dependant nodal admittance matrix according to the general rule
\begin{align}\label{eq: i=Yv}
    {\boldsymbol i}(s)={\bf Y}(s) {\boldsymbol v}(s).
\end{align}
Matrix ${\bf Y}(s)$ includes contributions from the network (lines), i.e., ${\bf Y}_{N}(s)$, and from the shunt elements\footnote{Shunt elements are any elements which tie the network to ground, and they can include loads, sources, and shunt network elements.}, i.e., ${\bf Y}_{S}(s)$. This latter part is diagonal. Written explicitly, 
\begin{align}\label{eq: Y_sum}
{\bf Y}(s)=\!\!\sum_{\{i,j\}\in {\mathcal E}}\!\!\!{\bf Y}_{N}^{(i,j)}(s)+\sum_{i\in {\mathcal V}}{\bf Y}_{S}^{(i)}(s),
\end{align}
where ${\bf Y}_{N}^{(i,j)}$ is the matrix associated with the network connection between buses $i$ and $j$, and ${\bf Y}_{S}^{(i)}$ is the matrix associated with shunt elements at bus $i$. For example, the simple $2$-bus network illustrated in Fig. \ref{fig: Closed_Loop_Network} can be decomposed and written as
\begin{subequations}\label{eq: Y_decomp_Toy}
\begin{align}
{\bf Y} & =\;\;\quad\quad{\bf Y}_{N}^{(1,2)}\quad\;\,+\;\quad\;{\bf Y}_{S}^{(1)}\quad+\quad\;\;{\bf Y}_{L}^{(2)}\\[-5pt] 
 & =\overbrace{\left[\begin{array}{cc}
Y_{l}\! & \!-Y_{l}\\
-Y_{l}\! & \!Y_{l}
\end{array}\right]}^{}+\overbrace{\left[\begin{array}{cc}
Y_{S} & 0\\
0 & 0
\end{array}\right]}^{}+\overbrace{\left[\begin{array}{cc}
0 & 0\\
0 & Y_{L}
\end{array}\right]}^{}.
\end{align}
\end{subequations}
{We note that the network admittance matrix ${\bf Y}_{N}(s)$ is essentially the dynamic generalization of the conventional admittance matrix used in power flow applications \cite{machowski2011power}. Thus, by including electromagnetic dynamics, the admittance of the line between buses $i$ and $j$ can be written as: 
\begin{equation}
    Y_{ij}(s)=\frac{1}{R_{ij}+s L_{ij}}
\end{equation}
where $R_{ij}$ and $L_{ij}$ are the resistance and the inductance of the line, respectively. Expressions for load effective admittances $Y_{L}$ are more complex, and will be provided in the next subsection.}
In the absence of external perturbations, \eqref{eq: i=Yv} reduces to
\begin{equation}\label{eq: 0=Yv}
    {\boldsymbol 0}={\bf Y}(s) {\boldsymbol v}(s),
\end{equation}
which will always be satisfied, since the network obeys Kirchhoff's laws. In general, (\ref{eq: 0=Yv}) may only be satisfied by nontrivial solutions of ${\boldsymbol v}(s)$ when $s=s_0$ is chosen such that ${\bf Y}(s_0)$ is a singular matrix.
\begin{definition}\label{eq: Def_Eig} ${\boldsymbol v}e^{s_0t}$, ${\boldsymbol v}\ne {\bf 0}$, is an \textbf{eigenmode} of ${\bf Y}(s)$ if
\begin{align}\label{eq: 0=Yvest}
    {\bf 0} = {\bf Y}(s_0){ {\boldsymbol v}}e^{s_0t}.
\end{align}
\end{definition}
In the following, both $s_0$ and ${\boldsymbol v}e^{s_0t}$ will be referred to as the same eigenmode. As a necessary consequence of (\ref{eq: 0=Yvest}), ${\rm det}[{\bf Y}(s_0)]=0$ if $s_0$ is an eigenmode. We now collect all values of $s_0$ which satisfy Definition \ref{eq: Def_Eig} and place them in the vector ${\boldsymbol s}_0$. The small-signal stability of system (\ref{eq: i=Yv}) can be gauged by computing the real parts of eigenmode vector ${\boldsymbol s}_0$, as stated by the following Lemma:
\begin{lemma}\label{lemma:eig_mode}
System (\ref{eq: 0=Yv}) is stable iff
\begin{align}\label{eq: real_ems}
{\rm Re}\{s_0\}<0,\,\forall s_0 \in {\boldsymbol s}_0.
\end{align}
\begin{proof}
In the absence of external perturbations, the solution to (\ref{eq: 0=Yv}) will be a scaled sum of its eigenmodes. The decay rate of each eigenmode is given by ${\rm Re}\{s_0\}$. If (\ref{eq: real_ems}) holds, then the solution will decay exponentially with time.
\end{proof}
\end{lemma}
\noindent {If (\ref{eq: real_ems}) is satisfied, a direct corollary of Lemma \ref{lemma:eig_mode} implies the following: after some external perturbations are applied to a microgrid described by (\ref{eq: i=Yv}), the system will converge exponentially back to the stable equilibrium.}

{Using the stability assessment framework of Lemma \ref{lemma:eig_mode}, we can state the central problem considered in this manuscript.
\begin{problem}[\textbf{Central Problem Statement}]\label{prob: CPS}
Assume a system ${\bf Y}(s)$, which satisfies (\ref{eq: 0=Yv}), is stable via Lemma \ref{lemma:eig_mode}. Components are then added to, or removed from, ${\bf Y}(s)$ to create a new system ${\bf Y}_n(s)$. How can ${\bf Y}_n(s)$ be ensured to satisfy Lemma \ref{lemma:eig_mode} without a centralized computation of its new eigenmodes?
\end{problem}
To solve the challenge posed by Problem \ref{prob: CPS}, this paper leverages decentralized techniques which are sufficient to guarantee that no eigenmodes can cross into the unstable Right Half Plane (RHP) when ${\bm Y}(s)$ is altered.}

\subsection{DC Load Model}
DC-DC power electronic converters are designed with a control system which properly mitigates input voltage variations in order to provide a fairly constant load voltage. A common way to represent a lossless converter is using a single-pole, double-throw (SPDT) switch with a duty cycle $D$. Due to the switch, which operates with a frequency on the order of tens of kilohertz, the inductor current and capacitor voltage vary throughout a given cycle with small amplitude around their average values even in steady state. Therefore, to properly account for the low frequency variations, the so-called small-ripple approximation is used, where the system is averaged over a given period and switching variations are neglected.

{There are many different types of converters that are used depending on particular applications. In this manuscript, we will focus on the buck and boost converters due to their prevalence; however, our methods and results can be easily generalized to any converter model. Below we describe in more details a buck converter. Expressions for a boost converter can be obtained in a similar way and are given in Appendix \ref{Boost_Converter}.} 

\begin{figure}
    \centering
    \includegraphics[width=1\linewidth]{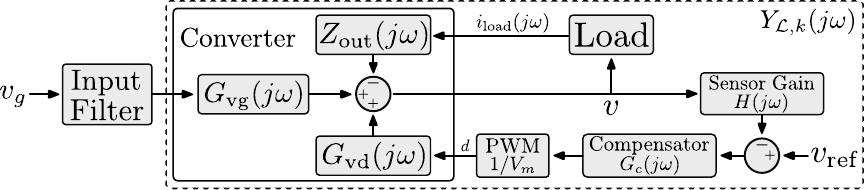}
    \caption{Definition of Small-Signal Load Model}
    \label{fig:Load_Model}
\end{figure}

\begin{figure}
    \centering
    \includegraphics[width=1\columnwidth]{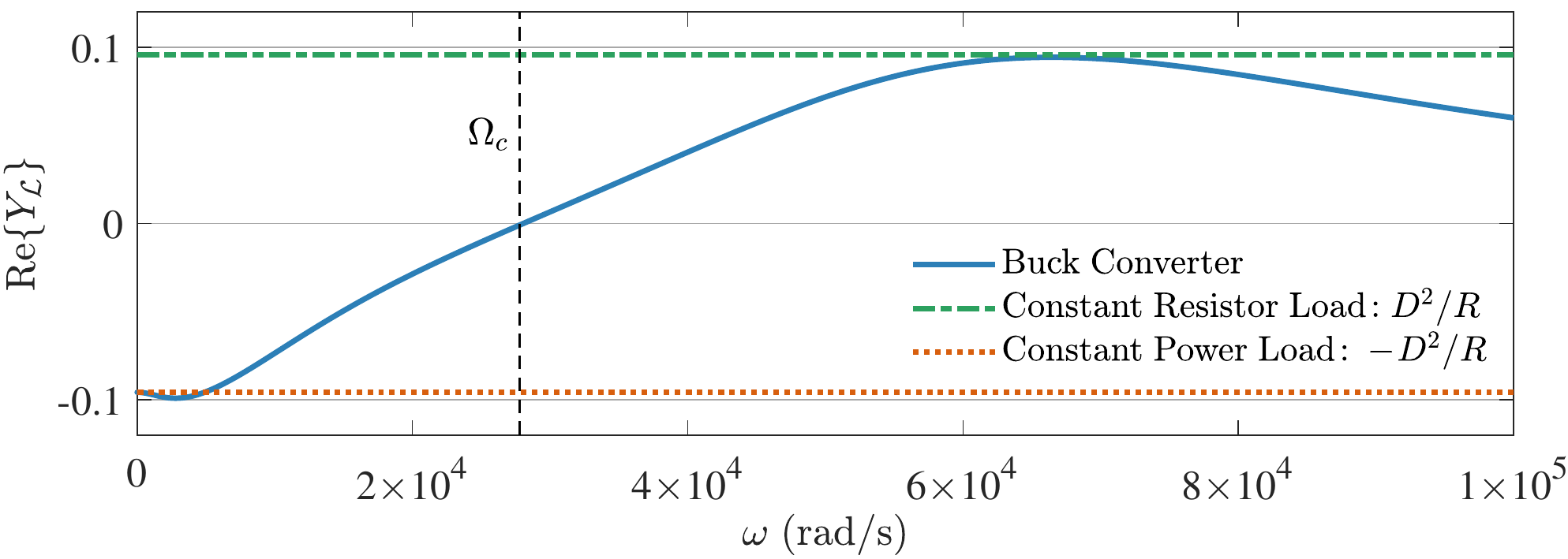}
    \caption{Shown are the real parts of the input admittance for a buck converter (blue), constant resistance load (green), and constant power load (orange).}
    \label{fig:BuckTF}
\end{figure}

The buck converter, consisting of an inductance $L$, capacitance $C$ and constant resistance load $R$, has a step-down conversion ratio $D$. Since averaging is done over many switching periods, an ideal DC transformer, rather than a switch, is used in the small-signal model. Further buck converter details are provided in \cite{erickson2007fundamentals}. To maintain constant voltage at its output, a converter relies on feedback to modify the duty cycle following any variations in the input voltage. In this work, we consider a buck converter interfaced with a constant resistance load on its output and a control loop to stabilize the output voltage. In this way, it is seen from the network as a constant power load in steady state. This model is represented in Fig. \ref{fig:Load_Model}. To characterize the small-signal behavior of the buck converter, we first examine the transfer function from the control input (duty cycle variation) $d(s)$ to the output voltage $v(s)$ \cite{erickson2007fundamentals}: 
\begin{equation}\label{eq: c2otf}
    G_{vd}(s) = \frac{v(s)}{d(s)} = G_{d0}\frac{\omega_0^2}{s^2+2\zeta\omega_0s+\omega_0^2}
\end{equation}
with $G_{d0} = V/D$, $\omega_0 = 1/\sqrt{LC}$ and $\zeta = \sqrt{L/C}/(2R)$. In the presence of voltage control, the loop gain of the system is defined as $T = G_{vd}(s)H(s)G_c(s)/V_m$, where $H$ is the sensor gain, $G_c(s)$ is the compensator transfer function and $V_m$ is the voltage of the pulse width modulation. The compensator is modeled as a simple lead-lag controller of the form 
\begin{equation}\label{eq: comp}
    G_c(s) = G_{c\infty}\frac{(1+\frac{\omega_L}{s})(1+\frac{s}{\omega_z})}{(1+\frac{s}{\omega_p})}.
\end{equation}
The input admittance transfer function $Y_{\mathcal{L}}(s)$ can be written
\begin{equation}\label{eq: BCL_TF}
    Y_{\mathcal{L}}(s) =\frac{i(s)}{v_g(s)} = \frac{1}{Z_N(s)}\frac{T(s)}{1+T(s)}+\frac{1}{Z_D(s)}\frac{1}{1+T(s)}
\end{equation}
where, for a buck converter,
\begin{itemize}
    \item $Z_N=\frac{-R}{D^2}$ is the converter impedance when the load voltage perturbations $v(s)$ are eliminated (i.e., $d(s)$ is infinitely fast and the load acts like constant power), and
    \item $Z_D=\frac{R}{D^2}\frac{1+sL/R+s^2LC}{1+sRC}$ is the converter impedance in the absence of control input $d(s)$ (i.e., the load acts like constant impedance).
\end{itemize}
In steady state, $Z_D=-Z_N$. This is shown quite clearly by the green and orange traces in Fig \ref{fig:BuckTF}, where above $\Omega_c$, $\text{Re}\{Y_\mathcal{L}(\omega)\}>0$, and the load behaves more like a passive impedance. In practice, converters are also equipped with an input filter to reject the high-frequency current components from converter switching actions. Improper input filter design can bring instability to an otherwise stable converter. 

\section{Decentralized Stability Certificates}\label{sec:genap}
{In this section we present the derivation of the \emph{network topology invariant} stability assessment method -- the central result of the manuscript. The method is based on a simple idea that if a certain dynamic system is being continuously transformed (by means of smoothly varying some of its parameters) from a stable state to an unstable one, then at least one of its eigenvalues crosses the imaginary axis during this transition. Indeed, all the eigenvalues of a stable system lie in the left-hand part of the complex plane, while for an unstable system at least one of the eigenvalues is in the right-hand part. When the system is being transformed continuously, the eigenvalues move in the complex plane, and those that will be in the RHP for the final state should cross the imaginary axis during the transition.} 

{The described property of the eigenvalues give an idea on how stability of a certain system can be certified. First, we need to somehow ``parameterize" the system in question so that by varying the parameter (or parameters) we can transform it to a state that we know is definitely stable. If now we make sure, that during the transformation none of the eigenvalues appear on the imaginary axis, we can certify that the system is stable. In order to give a better explanation of this concept, the next subsection considers a simple example. After that, a more rigorous mathematical formulation is given for a particular case of a DC network. Then, the decentralized stability assessment method is derived. }

\subsection{Parameterization Example}
{Suppose we have a dynamic SISO system with the input-to-output transfer function $h_0(s)$, i.e., $y(s)=h_0(s)u(s)$ (where $u(s)$ is the input, and $y(s)$ is the output). Suppose also, that we know that system is stable, i.e., all the poles of $h_0(s)$ are in the left-hand part of the complex plane. Now, we would like to consider the new system, where negative feedback with some gain $K$ is added, and study its stability. This new system input to output function is now:
\begin{equation}\label{eq:feedback}
    h_f(s) = \frac{h_0(s)}{1+K\cdot h_0(s)}
\end{equation}
where the subscript $f$ means ``feedback". We now choose to ``parameterize" our model with a homotopic scalar gain value $\alpha \in [0,\, 1]$, where $\alpha = 1$ corresponds to the feedback configuration \eqref{eq:feedback} and $\alpha=0$ corresponds to the stable ``open loop" system $h_0(s)$. A straightforward way to realize this would be to multiply the system feedback gain $K$ by $\alpha$\footnote{{More generally, $\alpha$ can be used to scale \textit{any} driver of instability in the microgrid network (e.g., line inductance $L$). In this paper, we focus exclusively on closed loop control as the instability driver.}}, so that we get
\begin{equation}\label{eq:feedbackpar}
    h_f(s) = \frac{h_0(s)}{1+(\alpha \cdot K)\cdot h_0(s)}.
\end{equation}}
Clearly, when $\alpha=0$, the closed feedback loop is broken. The primary feature of this parameterization scheme is that for $\alpha=0$, the system is ``definitely" stable, i.e., all of its eigenmodes are in the LHP (due to our initial assumption about $h_0(s)$). Thus, setting $\alpha=0$ is equivalent to setting feedback controller gain 0, so that the system operates in an open loop configuration. When $\alpha$ is scaled up to $1$, then the controller gain gradually scales up to its nominal value. As will be clarified, the scaling of $\alpha$ from $0$ to $1$ can be thought of as a homotopy procedure, since it represents a smoothly varying parameterization of the system eigenmodes.

{The simple example \eqref{eq:feedbackpar} is just one of the infinitely many possible parametrization schemes. In this manuscript we will be using this type of parametrization for stability assessment of DC microgrids. Thus, the effective admittance matrix of a network in question takes the from ${\bf Y}(s,\alpha)$, where we assume, that when $\alpha=0$ the system is definitely stable. For instance, if following the same logic as in example \eqref{eq:feedbackpar}, we use $\alpha$ to scale the feedback gains of all the converters in our network, then $\alpha=0$ means that all the converters are operating in an open-loop configuration, essentially representing some effective $RLC$ circuits (i.e. converters operating with constant duty cycles). Such a system is small-signal stable.}

\subsection{{Stability Assessment via Admittance Matrix Singularity}}
Using the system parametrization illustrated in the previous subsection, matrix ${\bf Y}(s,\alpha)$ is used to denote the previously derived small-signal network model ${\bf Y}(s)$ which has been parameterized with $\alpha$. The general strategy for proving the stability of the system ${\bf Y}(s,\alpha)$ is based on the well-known ``zero exclusion principle'' and is closely related to the $\mu$-analysis, as is shown in the following stability lemma.
\begin{lemma}\label{lemma: det_nz}
Assuming ${\bf Y}(s,\alpha)$ is stable for $\alpha=0$, then ${\bf Y}(s,\alpha)$ is also stable for $\alpha=1$ if the determinant satisfies
\begin{align}\label{eq: det_Y}
{\rm det}[{\bf Y}(s=j\omega,\alpha)]\ne0,\;\forall\omega\ge0,\;\forall \alpha\in[0,\,1].
\end{align}
\begin{proof}
For $\alpha=0$, the real part of all eigenmodes will lie in the LHP. For the system to go unstable, an eigenmode must cross the imaginary axis. At the point of crossing, $\exists \, \alpha,\,\omega \ni {\rm det}[{\bf Y}(s=j\omega,\alpha)]=0$, since one of the eigenmodes will necessarily have ${\rm Re}\{s_0\}=0$. If this does not occur, then the system must remain stable.
\end{proof}
\end{lemma}
\noindent To generalize this lemma, we introduce the rotational function 
\begin{align}
{\mathcal D}(\omega,\alpha) = e^{j\phi(\omega,\alpha)}
\end{align}
whose magnitude is unity and whose phase function $\phi(\omega,\alpha)$ depends on both frequency and the parameterization value $\alpha$. {It is important that this function never turns to zero magnitude for any value of $\phi(\omega,\alpha)$.}
\begin{remark}
The results of lemma \ref{lemma: det_nz} remain valid if ${\bf Y}(s,\alpha)$ is multiplied by any nonzero function. Thus, we can write the generalization of condition (\ref{eq: det_Y}) as
\begin{align}\label{eq: det_DY}
{\rm det}[{\mathcal D}(\omega,\alpha){\bf Y}(s=j\omega,\alpha)]\ne0,\;\forall\omega\ge0,\;\forall \alpha\in[0,\,1].
\end{align}
\end{remark}
While (\ref{eq: det_DY}) can guarantee the stability of the system ${\bf Y}(s,\alpha)$, it is still a centralized criterion, i.e., there is no way to relate the determinants of the subsystems to the determinant of the full system. Rather than requiring a nonzero determinant, a somewhat more conservative condition is to require the positive definiteness of the matrix in question over the same range of parameters. While being conservative, positive definiteness has a very useful additive property, i.e., if the matrices in a set are all positive definite, then so is their sum. {We make the following remark, which clarifies the connection between matrix singularity and positive definiteness:
\begin{remark}\label{rem_pd}
Consider matrix ${\bm A}\in{\mathbb C}^{n \times n}$, which is generally non-Hermitian. The following implications hold $\forall {\bm A}$~\cite{Horn:1990}:
\begin{itemize}
    \item ${\rm rank}\{\bm{A}\}\ne n\;\;\;\Rightarrow\;\;\;\bm{A}+\bm{A}^{\dagger}\nsucc0$
    \item $\bm{A}+\bm{A}^{\dagger}\succ0\;\;\;\;\;\Rightarrow\;\;\;{\rm rank}\{\bm{A}\}=n$.
\end{itemize}
\end{remark}}

{\noindent Remark \ref{rem_pd} will be used in the proof of the following Lemma.}
\begin{lemma}\label{lemma: PD_Mats}
Assume network ${\bf Y}\equiv{\bf Y}(s,\alpha)$ is stable for $\alpha=0$. Set $s=j\omega$. If there exists ${\mathcal D}\equiv e^{j\phi(\omega,\alpha)}$ for which
\begin{align}\label{eq: PD_mats}
{\mathcal D}{\bf Y}+\left({\mathcal D}{\bf Y}\right)^{\dagger}\succ0,\;\forall\omega\ge0,\;\forall \alpha\in[0,\,1],
\end{align}
then ${\bf Y}(s,\alpha)$ is stable for all gain values up to $\alpha=1$.
\begin{proof}
Assume matrix $\bf Y$ becomes singular for some allowed values of $\omega$, $\alpha$. Then ${\mathcal D}{\bf Y}$ will also be singular, meaning ${\mathcal D}{\bf Y}+\left({\mathcal D}{\bf Y}\right)^{\dagger}$ cannot be positive definite, {by Remark \ref{rem_pd}}. If (\ref{eq: PD_mats}) is satisfied, therefore, $\bf Y$ cannot become singular, and Lemma \ref{lemma: det_nz} holds.
\end{proof}
\end{lemma}

\subsection{Decentralized Stability Certificate}
As written, (\ref{eq: PD_mats}) is still a fully centralized stability certificate, because ${\bf Y}$ is the full system matrix. The certificate can be converted to a decentralized one, though, by writing ${\bf Y}$ as a summation of its network and shunt components, as in (\ref{eq: Y_sum}). To simplify notation, we introduce set ${\mathcal C}={\mathcal E}\cup{\mathcal V}$; it is defined to be the set of indices associated with all dynamical elements in the network. Furthermore, $Y_{k} \equiv Y_{k}(s,\alpha),\;k\in{\mathcal C}$ refers to the $k^{\rm th}$ complex \textit{{scalar}} transfer function element in the system (line, load, or source). {We are now ready to prove the following theorem, which is one of the main results of the manuscript.}
\begin{theorem}\label{theorem: PD Mats Decentralized}
Assume network ${\bf Y}\equiv{\bf Y}(s,\alpha)$ of (\ref{eq: 0=Yv}) is stable for $\alpha=0$. Set $s=j\omega$. \textbf{If} there exists ${\mathcal D}\equiv e^{j\phi(\omega,\alpha)}$ for which
\begin{align}\label{eq: DYpDY}
{\rm Re}\{{\mathcal D}Y_{k}\}>0,\;\forall k\in\mathcal{C},\;\forall\omega\ge0,\;\forall \alpha\in[0,\,1],
\end{align}
then ${\bf Y}(s,\alpha)$ is stable for all gain values up to $\alpha=1$.
\begin{proof}
Using (\ref{eq: Y_sum}), we break system $D{\bf Y}$ into
\begin{align}\label{eq: DY = }
{\mathcal D}{\bf Y}=\underbrace{\textstyle\sum_{i,j\in\mathcal{E}}{\mathcal D}{\bf Y}_{N}^{(i,j)}}_{{\bm A}}+\underbrace{\textstyle\sum_{i\in\mathcal{V}} {\mathcal D}{\bf Y}_{S}^{(i)}}_{{\bm B}},
\end{align}
{where ${\bf Y}_{N}^{(i,j)},{\bf Y}_{S}^{(i)},{\bm A},{\bm B}\!\in\!{\mathbb C}^{n\times n}$.} Assume (\ref{eq: DYpDY}) holds true. By construction, {Hermitian} matrix ${\bm A}_H={\bm A}+{\bm A}^{\dagger}$ is positive semi-definite (PSD) with a single $\lambda=0$ eigenvalue. The eigenvector associated with this zero eigenvalue is equal to ${\bf e}=[1, 1, \ldots, 1]^{\top}$. Matrix ${\bm B}_H={\bm B}+{\bm B}^{\dagger}$ is also at a PSD matrix, but it is also diagonal, so ${\bf e}^{\dagger}{\bm B}_H{\bf e}>0$. Thus, (\ref{eq: DYpDY}) implies positive definiteness of ${\mathcal D}{\bf Y}+\left({\mathcal D}{\bf Y}\right)^{\dagger}$, which further implies the {nonsingularity of ${\bf Y}$ by Remark \ref{rem_pd}}, and thus the stability of ${\bf Y}(s,\alpha)$ up to $\alpha=1$.
\end{proof}
\end{theorem}
\noindent As a helpful simplification, we may use polar coordinates to write $\theta_k\equiv\angle Y_{k}(\omega,\alpha)$ and $|Y_{k}|\equiv|Y_{k}(\omega,\alpha)|$. Thus, the positivity condition in (\ref{eq: DYpDY}) can be stated as
\begin{subequations}\label{eq: DY_simp}
\begin{align}
0 & < {\mathcal D}Y_{k}+\left({\mathcal D}Y_{k}\right)^{\dagger}\\
 & =2|Y_{k}|{\rm Re}\{e^{j\left(\phi(\omega,\alpha)+\theta_k\right)}\}.
\end{align}
\end{subequations}
Using (\ref{eq: DY_simp}), we have the following corollary.
\begin{corollary}\label{corollary: phase_simp}
\textbf{If} there exists $\phi\equiv\phi(\omega,\alpha)$ for which
\begin{align}\label{eq: phase_cond}
-\frac{\pi}{2}\!<\!\phi+\theta_k\!<\!\frac{\pi}{2},\,\forall k\!\in\!\mathcal{C},\,\forall\omega\ge0,\,\forall \alpha\!\in\![0,1],
\end{align}
and ${\bf Y}(s,\alpha=0)$ is stable, \textbf{then} ${\bf Y}(s,\alpha\!=\!1)$ is also stable.
\end{corollary}

\subsection{The Problem of Negative Admittance at Zero Frequency}
Assuming the network model is constructed using physical laws and physically realizable controllers, the admittance matrix ${\bf Y}(s=j0,\alpha)$ will be a purely real matrix. Therefore, the angle $\theta_k$ associated with all elements in this matrix will have an angle of either $\theta_k=0^{\circ}$ or $\theta_k=180^{\circ}$. If any shunt element, such as a constant power load, has $\theta_k=180^{\circ}$ at $s=j0$, then condition (\ref{eq: phase_cond}) cannot be satisfied at this frequency, since any network line will necessarily have $\theta_k=0^{\circ}$. Mathematically, this limiting case of $s=j0$ cannot be ignored, since a negative real eigenmode (i.e., system pole) may potentially travel on the real axis and cross the imaginary axis at $s=j0$. Therefore, network singularity at zero frequency must be considered.

\begin{definition}
System ${\bf Y}(s,\alpha)$ is said to be \textbf{zero-frequency stable} if ${\rm det}[{\bf Y}(s\!=\!j0,\alpha)]\ne0,\,\forall\alpha\in[0,1]$. That is, no real eigenmode crosses into the RHP as $\alpha$ is scaled from $0$ to $1$.
\end{definition}
We note, however, that ${\rm det}[{\bf Y}(s=j0,\alpha)]=0$ will occur at the point of maximum steady-state network loadability. Therefore, if this determinant is close to 0, then the system is operating near the tip of the so-called ``nose curve". Practical systems are engineered to operate far from this nose curve. We therefore offer the following assumption, which entirely precludes the possibility of a system encountering zero-frequency instability.
\begin{assumption}\label{asmpt: SS_stable}
System (\ref{eq: 0=Yv}) is assumed to operate far from the steady state tip of the nose curve and is therefore zero-frequency stable.
\end{assumption}
\noindent With the addition of Assumption \ref{asmpt: SS_stable}, we update Corollary \ref{corollary: phase_simp}.
\begin{tcolorbox}
\begin{corollary}\label{corollary: phase_simp2}
\textbf{If} there exists $\phi\equiv\phi(\omega,\alpha)$ for which
\begin{align}\label{eq: phase_cond2}
-\frac{\pi}{2}\!<\!\phi+\theta_k\!<\!\frac{\pi}{2},\forall k\!\in\!\mathcal{C},\,\forall\omega>0,\,\forall \alpha\!\in\![0,1],
\end{align}
and Assumption \ref{asmpt: SS_stable} holds, and ${\bf Y}(s,\alpha=0)$ is stable, \textbf{then} ${\bf Y}(s,\alpha)$ is also stable $\forall \alpha \in [0,1]$.
\end{corollary}
\end{tcolorbox}
Per (\ref{eq: phase_cond2}), the phase condition must be satisfied for all \textit{positive} $\omega$, rather than all non-negative $\omega$, as in (\ref{eq: phase_cond}). {The microgrid model ${\bm Y}(s,\alpha)$ can easily include any sort of local, primary control. To ensure the system will remain stable under a secondary control scheme (i.e., centralized set point reassignment), however, Corollary \ref{corollary: phase_simp2} must be additionally satisfied across the full set of potential set point updates. We further note that Corollary \ref{corollary: phase_simp2} is invariant to not only the topology of the network, but also to the \textit{size} of the network. That is, the small-signal stability condition will still hold, even for an arbitrarily sized network, if Corollary \ref{corollary: phase_simp2} stays true.}

\subsection{Simple $2$-bus Example}
\begin{figure}
\centering
\includegraphics[width=1\columnwidth]{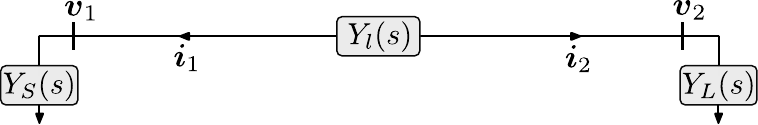}
\caption{\label{fig: Closed_Loop_Network} Simple 2 bus DC microgrid network. The source, load, and line dynamics are represented by $Y_S(s)$, $Y_L(s)$, and $Y_l(s)$, respectively.}
\end{figure}

As an illustrative example, we consider the network in Fig. \ref{fig: Closed_Loop_Network}. We define the open-loop load dynamics via
\begin{align} \label{eq: h1(s)}
h_0(s) & =\frac{1}{s^{2}+2\zeta\omega_{n}s+\omega_{n}^{2}}.
\end{align}
A PI controller is then added in a negative feedback loop to yield the resulting admittance function
\begin{align}\label{eq: Y(s,K)}
Y_L(s) & =\frac{h_0(s)}{K(1+\tfrac{1}{s\tau_i}) h_0(s)+1}.
\end{align}
The source is considered to be an ideal voltage source\footnote{Ideal voltage sources act as direct paths to ground in a small-signal admittance model.} connected through a short line. Thus, the admittances associated with the source and the network line are respectively given by
\begin{align}
Y_{S}(s) & =\frac{1/R_{s}}{1+s\tau}, \quad Y_{l}(s)  =\frac{1/R_{l}}{1+s\tau},
\end{align}
where, for simplicity, the time constant $\tau$ is assumed equal in both admittances. We consider the stability of this system in two ways: first, using the non-conservative centralized approach of Lemma \ref{lemma: det_nz}, and second, using the decentralized approach proposed by Corollary \ref{corollary: phase_simp2}.

\subsubsection{Centralized Stability Approach}
In the centralized approach of Lemma \ref{lemma: det_nz}, we may leverage the full, centralized network model. {In this case we can set $\alpha=1$ and directly determine the allowed range of the values for the controller gain $K$.} The determinant of the system admittance matrix (\ref{eq: Y_decomp_Toy}) is
\begin{align}
{\rm det}[{\bf Y}] & =\left(Y_{S}+Y_{l}\right)\left(Y_{L}+Y_{l}\right)-Y_{l}^{2}.\label{eq: det_0}
\end{align}
Satisfying $\left(Y_{S}+Y_{l}\right)\left(Y_{L}+Y_{l}\right)-Y_{l}^{2}=0$ simplifies to
\begin{align}\label{eq: s_cube_eq}
s^{3}+\left(\tau R+2\zeta\omega_{n}\right)s^{2}+\left(\omega_{n}^{2}+K+R\right)s+K/\tau_{i}=0,
\end{align}
where $R \equiv R_s+R_l$. By Routh-Hurwitz, an eigenmode will become unstable when $K>\tau_{i}(\tau R+2\zeta\omega_{n})(\omega_{n}^{2}+K+R)$. This simplifies to
\begin{align}\label{eq: K_unstable}
K>\frac{\tau_{i}\left(\tau R+2\zeta\omega_{n}\right)\left(\omega_{n}^{2}+R\right)}{1-\tau_{i}\left(\tau R+2\zeta\omega_{n}\right)}\equiv\hat{K}
\end{align}
assuming $1>\tau_{i}\left(\tau R+2\zeta\omega_{n}\right)$. When (\ref{eq: K_unstable}) is satisfied as an equality, the complex roots of (\ref{eq: s_cube_eq}) are
\begin{align}
s & =0\pm j\sqrt{(\omega_{n}^{2}+R)/(R\tau\tau_{i}+2\tau_{i}\zeta\omega_{n}-1)}\equiv j\hat{\omega}.
\end{align}
Since ${\rm det}[{\bf Y}(s=j{\hat \omega})]=0$ for $K={\hat K}$, this is the gain value for which stability is lost, and $\hat \omega$ is the frequency of the system mode that becomes unstable. When applying the numerical values $\omega_n=2\pi$, $\zeta = 0.1$, $\tau = 0.1$, and $\tau_i=0.25$, we find ${\hat K}=20.77$ is the maximum tolerable gain value {for the particular system of Fig.~\ref{fig: Closed_Loop_Network}}.

\subsubsection{Decentralized Stability Approach}
\begin{figure}
\centering
\includegraphics[width=1\columnwidth]{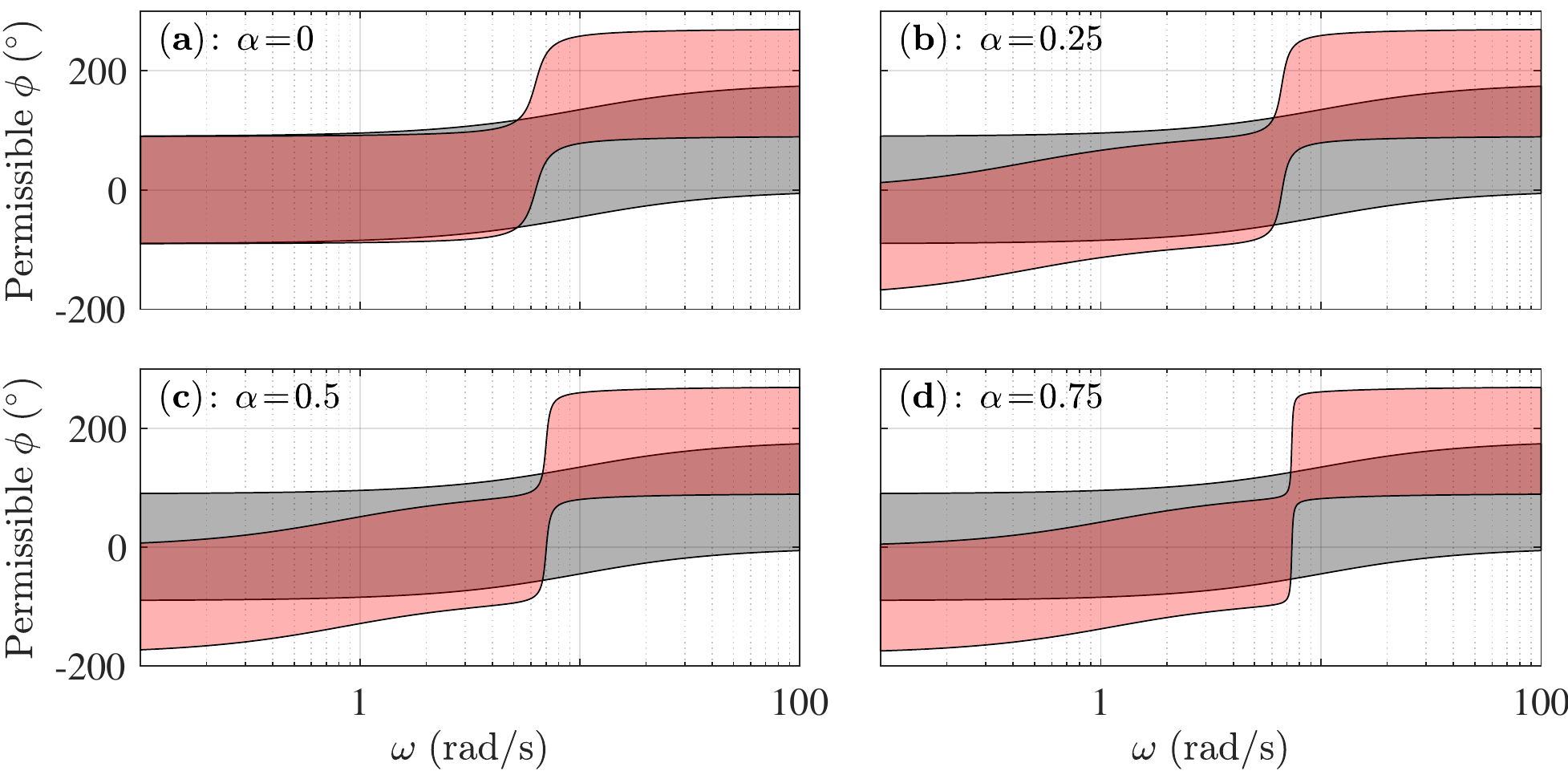}
\caption{\label{fig: Example_Plots4} The {intersection of the shaded sectors} are the values that function $\phi(\omega,\alpha)$ can take, according to inequality (\ref{eq: phase_cond2}), for a continuum of frequency and controller gain values. The grey sector corresponds to the line phase function (\ref{eq: line_phase_example}), while the red sector corresponds to the load phase function (\ref{eq: shunt_phase}). {Across all four panels, there is no loss of overlap between these shaded regions, indicating the existence of function $\phi(\omega,\alpha)$ for up to $\alpha=0.75$.}}
\end{figure}

\begin{figure}
\centering
\includegraphics[width=1\columnwidth]{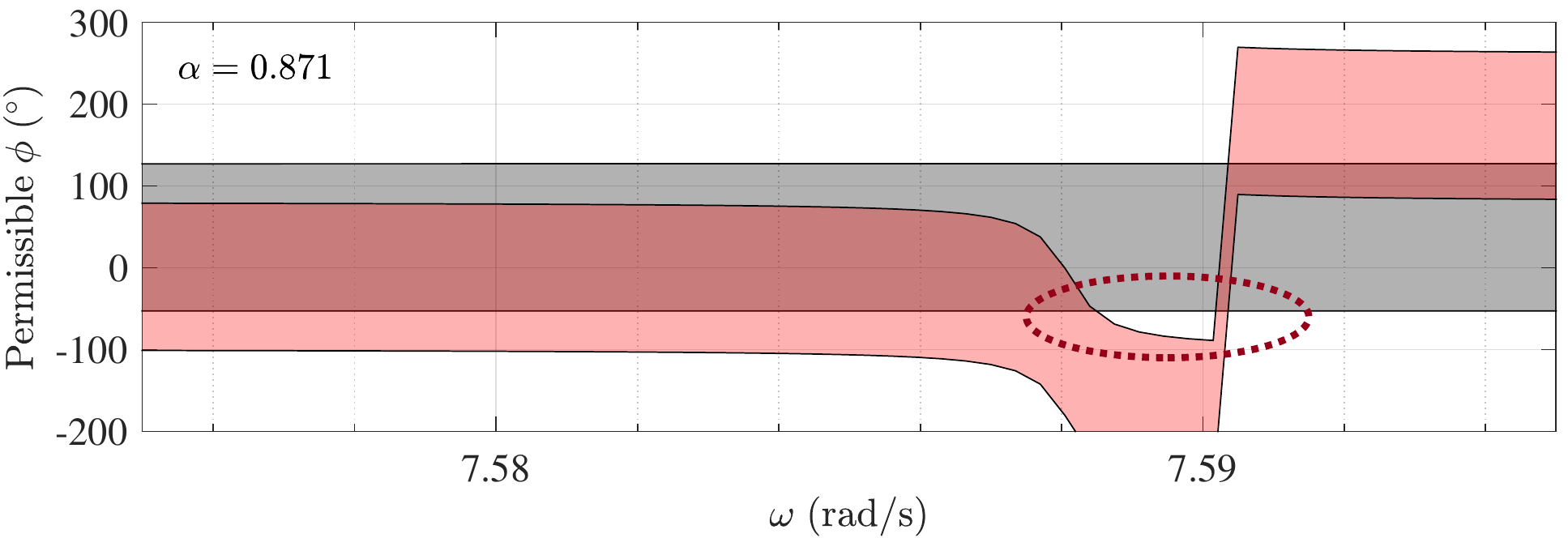}
\caption{\label{fig: Example_Plots1} {The intersection of the shaded sectors are the values that function $\phi(\omega,\alpha)$ can take, according to inequality (\ref{eq: phase_cond2}). Loss of continuous overlap between these shaded regions indicates the non-existence of function $\phi(\omega,\alpha)$; this overlap is clearly lost at $\omega\approx 7.59$ when $\alpha=0.871$.}}
\end{figure}

In the decentralized approach, we consider positive definiteness of the network elements independently via condition (\ref{eq: phase_cond2}). To apply this corollary, we parameterize the load by setting $K\rightarrow\alpha\cdot{\hat K}$ (we keep ${\hat K}=20.77$), so that for $\alpha = 1$, the system will definitely become marginally stable. Next, we independently define the phase functions $\theta_l(\omega,\alpha)$ and $\theta_L(\omega,\alpha)$ associated with parameterized admittance functions $Y_l(\omega,\alpha)$\footnote{Since they have identical time constants, the line and the source phase functions are identical. Thus, only one of them is defined.} and $Y_L(\omega,\alpha)$:
\begin{align}
\theta_l(\omega,\alpha)&=\tan^{-1}\left(-\omega\tau\right)\label{eq: line_phase_example}\\
\theta_L(\omega,\alpha)&=\tan^{-1}\biggl(\frac{\alpha\cdot{\hat K}/\tau_{i}-2\zeta\omega_{n}\omega^{2}}{\omega(\omega_{n}^{2}+\alpha\cdot{\hat K})-\omega^{3}}\biggr).\label{eq: shunt_phase}
\end{align}
Next, we plug these phase functions into inequality (\ref{eq: phase_cond}). We then plot the range of values which the unknown function $\phi(\omega,\alpha)$ can permissibly take, for each element, as \textit{continuous} functions of frequency, and as the homotopy parameter $\alpha$ is scaled up. {The results are shown in Figs. \ref{fig: Example_Plots4} and \ref{fig: Example_Plots1}, where the latter figure shows the magnified region where conditions of Corollary \ref{corollary: phase_simp2} are not satisfied. In these figures, 
\begin{itemize}
    \item the grey sector represents $-\theta_l(\omega,\alpha)\pm 90^{\circ}$,
    \item the red sector represents $-\theta_L(\omega,\alpha)\pm 90^{\circ}$,
    \item and their intersection represents the existence of $\phi$ in (\ref{eq: phase_cond2}).
\end{itemize}
If the gray and the red sector overlap for a certain value of frequency and $\alpha$, it means, that both $\theta_l$ and $\theta_L$ can be brought in the region $(-\pi/2,\pi/2)$ by properly choosing for both of them \textit{the same} rotation factor $\phi$. However, if the grey and red regions do not overlap at least for some frequencies and $\alpha$, as is the case in the Fig.~\ref{fig: Example_Plots1}, there is no rotation factor $\phi$ that can bring both $\theta_l$ and $\theta_L$ to the $(-\pi/2,\pi/2)$ region, so that condition \eqref{eq: phase_cond2} can not be satisfied.}

{In panels $({\bf a})$-$({\bf d})$ of Fig. \ref{fig: Example_Plots4}, there exists continuous overlap between the grey and red sectors, meaning stability of the system can be certified for up to $\alpha=0.75$. Fig. \ref{fig: Example_Plots1}, however, shows the sectors when $\alpha$ is increased up to $\alpha=0.871$. In this figure,} it can clearly be seen that the inequality ``breaks" at $\omega\approx7.59$. This corresponds to a maximum controller gain of $K\approx18.1$. When the sectors lose continuous overlap, the inequality is necessarily violated, and stability of the interconnection can no longer be guaranteed (even though the system damping ratio is still $0.8\%$). Although the result of $K\approx18.1$ is $12.9\%$ conservative, according to the proofs of the previous section, this gain value guarantees stability not only for the particular circuit in Fig. \ref{fig: Closed_Loop_Network}, {but also for an arbitrary interconnection of an arbitrary number of loads $Y_L(s)$ to an arbitrary number of sources $Y_S(s)$ using lines $Y_l(s)$. Thus, rather than providing a stability certificate for the specific configuration of Fig.~\ref{fig: Closed_Loop_Network}, we showed those components to be compatible (in the small-signal sense) with a stability condition which is network topology invariant.}

\section{Stability of DC Microgrids} \label{sec:num}
In this section, we test {various microgrid systems} consisting of three primary components: (i) stiff voltage sources, (ii) RL lines, and (iii) power electronics converters feeding constant resistance loads. In the first subsection, we explicitly parameterize the buck converter model presented in Fig. \ref{fig:Load_Model}. Then, we analyze the small-signal stability of the 10-bus radial network given in panel (\textbf{a}) of Fig. \ref{fig: MG_Power} under two different controller configurations. {Next, we test the stability result under a topology change, i.e., on a meshed 8-bus microgrid system. Subsequently, we analyze the effects of including boost converters in such a network, and then we consider the effects that time delay may have on our methods. Finally, we offer direct comparisons between our stability assessment method and other state-of-the-art methods.}

\subsection{Buck Converter Parameterization}
In the following tests, the loads are modeled as constant resistors interfaced with the network through regulated converters that are tuned to keep the output voltage constant. {In this subsection, we explicitly parameterize the buck converter model presented in Fig. \ref{fig:Load_Model} (the process for boost converter parameterization is identical)}. The nominal buck converter and controller parameters have been drawn from \cite{erickson2007fundamentals} and are defined in Table \ref{tab:sim_param_CC}. The buck converter's transfer function, with the additional grounded filter capacitor from Fig. \ref{fig:Load_Model}, can be parameterized by scaling the compensator gain $G_{c\infty}$ from (\ref{eq: comp}). Thus,
\begin{align}\label{eq: comp_gain}
\ensuremath{G_{c}(s,\alpha)}\equiv\alpha\cdot G_{c}(s)
\end{align}
and the full transfer function can be written as
\begin{align}\label{eq: TF_buck_Cf}
Y_{\mathcal{L}_c}(&s,\alpha)\triangleq \frac{V_{m}D^{2}}{V_{m}+G_{c}(s,\alpha)G_{vd}(s)H}\cdot\nonumber\\
\biggl(&\frac{1+sRC}{R+sL+s^{2}RLC} -G_{c}(s,\alpha)\frac{G_{vd}(s)H}{RV_{m}}\biggr)+C_{f}s,
\end{align}
where $C_{f}$ represents the parallel filter capacitor. At non-zero frequencies, when $\alpha\rightarrow0$, (\ref{eq: TF_buck_Cf}) relaxes to
\begin{align}
Y_{{\mathcal{L},c}}(s,\alpha\rightarrow0)=D^{2}\left(\frac{1+sRC}{R+sL+s^{2}RLC}\right)+C_{f}s,
\end{align}
which is effectively a passive RLC transfer function. Thus, the associated parameterized system will be \textit{definitely stable} when $\alpha=0$. When $s\rightarrow j0$, the admittance relaxes to
\begin{align}\label{eq: YL_SS}
Y_{\mathcal{L},c}(s\rightarrow j0,\alpha)= & -D^{2}/R,
\end{align}
as would be expected, {since a converter with output voltage regulation represents a constant power load at low frequencies}. Since the zero frequency response of the load is negative, the conditions from Corollary \ref{corollary: phase_simp2} must be satisfied in order to certify the stability of {a microgrid containing these elements.} The phase response of (\ref{eq: TF_buck_Cf}) is analytically cumbersome, but it can be numerically computed as
\begin{align}\label{eq: buck_phase}
\theta_{\mathcal{L}}(\omega,\alpha)=\tan^{-1}\left(\frac{{\rm Im}\left\{ Y_{\mathcal{L}}(s=j\omega,\alpha)\right\} }{{\rm Re}\left\{ Y_{\mathcal{L}}(s=j\omega,\alpha)\right\} }\right).
\end{align}

\subsection{Stability Test{: 10-Bus Radial Network}}
{We now gauge the stability of the radial network in panel (\textbf{a}) of Fig. \ref{fig: MG_Power}. We extend this system into a 10-bus network (with 9 radial loads and a single source).} The details associated with initializing this system (and its meshed equivalent) are located in Appendix \ref{App_Init}. In this radial system, the generators are assumed to be stiff voltage sources. Accordingly, their only dynamics are captured by the interconnecting line. The phase response of the RL lines (including the source lines) is given according to
\begin{align}\label{eq: line_phase}
\theta_{Z}(\omega,\alpha)=\tan^{-1}(-\tau\omega),
\end{align}
where $\tau=L/R$. {In this network, all loads are interfaced through buck converters with a lead-lag voltage compensator. All the parameters are present in Table \ref{tab:sim_param_CC}.} We will preform stability assessment using an exact, centralized approach (i.e., root locus, {which we refer to more specifically as \textit{eigenmode} locus}), and via the decentralized approach of Corollary \ref{corollary: phase_simp2}.

\begin{table}
    \centering
    \captionsetup{justification=centering}
    \caption{Converter, Controller, and Network Parameter Values {for the Numerical (NUM) and the Real-Time Simulation (RTS) Tests}}
    \label{tab:sim_param_CC}
    \begin{tabular}{|c|c|c|c|c|}
        \hline
          & \,\textbf{Parameter}\, & \;\;\;\;\textbf{Description}\;\;\;\; & \;\;{\textbf{NUM}}\;\;& \;\;\;{\textbf{RTS}}\;\;\; \\ \hline
          \multirow{5}{*}{\rotatebox[origin = c]{90}{\;\;\textbf{Converter}}}& $V$ & Input Voltage & $28\sim30$ V & {$28\sim30$ V}\\[-1.5pt]
          &$R$& Load (Mesh) & 3 $\Omega$ & {$3$ $\Omega$}\\[-1.5pt]
          &$R$ & Load (Radial) & 3 $\Omega$ & {$5\sim 36$} \\[-1.5pt]
          &$C$& Capacitance & 500 $\mu$F & {50 mF}\\[-1.5pt]
          &$L$& Inductance & 50 $\mu$H & {5 mH}\\[-1.5pt]
          &$D$&Duty Cycle& 0.536 & {$\sim0.53$}\\ \hline
          \multirow{6}{*}{\rotatebox[origin = c]{90}{\;\textbf{Controller}}}
          &$G_{c\infty}$& Midband Gain &3.7 &{3.7} \\[-1.5pt]
          &$w_L$& Lead Zero & $2\pi$500 Hz & {$2\pi$5 Hz}\\[-1.5pt]
          &$w_z$& Trailing Zero & $2\pi$1700 Hz & {$2\pi$17 Hz}\\[-1.5pt]
          &$w_p$& Pole & $2\pi$14.5 kHz & {$2\pi$145 Hz} \\[-1.5pt]
          &$H$& Sensor Gain & 1/3 & {1/3}\\[-1.5pt]
          &$V_m$& Voltage of PWM & 4 & {4}\\ \hline
          \multirow{4}{*}{\rotatebox[origin = c]{90}{\,\textbf{Network}}}
          &$C_f$& Filter Capacitor & 500 $\mu$F & {50 mF}\\[-1.5pt]
          &$l_{ij}$& Line Length & 0.1$\sim$1 km & {$0.1\sim$1 km}\\[-1.5pt]
          & $r$ & Line Resistance & 0.1 $\Omega$/km & {0.1 $\Omega$/km}\\[-1.5pt]
          &$\tau$& Line Time Constant & $1$ ms & {$100$ ms} \\[-1.5pt]
          &$V_s$& Source Voltage & 30 V & {30 V}\\ \hline
    \end{tabular}
\end{table}

\begin{figure}
    \centering
    \includegraphics[width=1\linewidth]{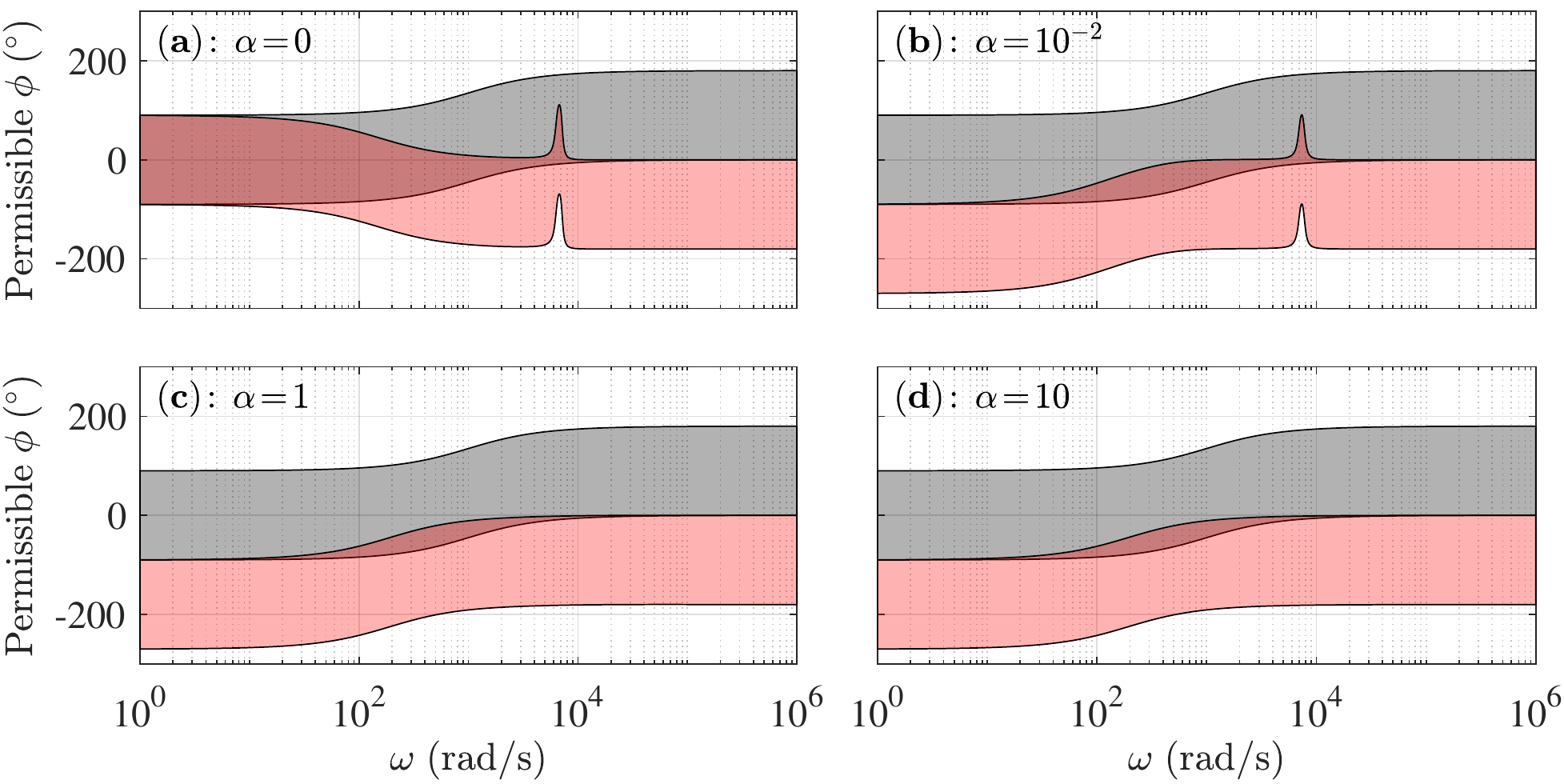}
    \caption{{Line and Buck Converter (Lead/Lag Compensator) Sector Plots.} Plotted are the values that $\phi(\omega,\alpha)$ can take, according to (\ref{eq: phase_cond2}), for a continuum of frequency and gain values. The grey sector corresponds to the line phase function (\ref{eq: line_phase}), while the red sector corresponds to the load phase function (\ref{eq: buck_phase}), where $G_c(s)$ is a lead-lag controller.}
    \label{fig:BuckLL_Curves}
\end{figure}

{In this example, lead-lag controllers are used as the compensators in each buck converter.} Specifically, $G_c(s)$ is the lead-lag controller specified in (\ref{eq: comp}). According to~\cite{erickson2007fundamentals}, the stability of (\ref{eq: BCL_TF}) is primarily controlled by the roots of $1+T(s)$. With the parameters from Table \ref{tab:sim_param_CC}, the lead-lag filter gives the bode plot of $1+T(s)$ a phase margin of $\phi_m=45.6^{\circ}$ and a gain margin of $g_m=\infty$. Using this controller, we first apply the decentralized approach of Corollary \ref{corollary: phase_simp2} by analyzing the phase functions (\ref{eq: buck_phase}) and (\ref{eq: line_phase}). These results are shown in Fig. \ref{fig:BuckLL_Curves}, where the phase functions are plotted from $\alpha=0$ up to $\alpha=10$ (i.e., scaling the gain $G_{c\infty}$ to 10 times its nominal value). As can be seen, there exists continuous overlap between the sectors; this is true for all loads in the system (results not shown), despite the small differences in equilibrium input voltages for the loads. Since the network is ``definitely stable" for $\alpha=0$, Corollary \ref{corollary: phase_simp2} is satisfied, and the system is \textit{guaranteed stable under arbitrary interconnection}, assuming input voltages remain sufficiently close to those tested (i.e., within a few volts).

\begin{figure}
    \centering
    \includegraphics[width=1\linewidth]{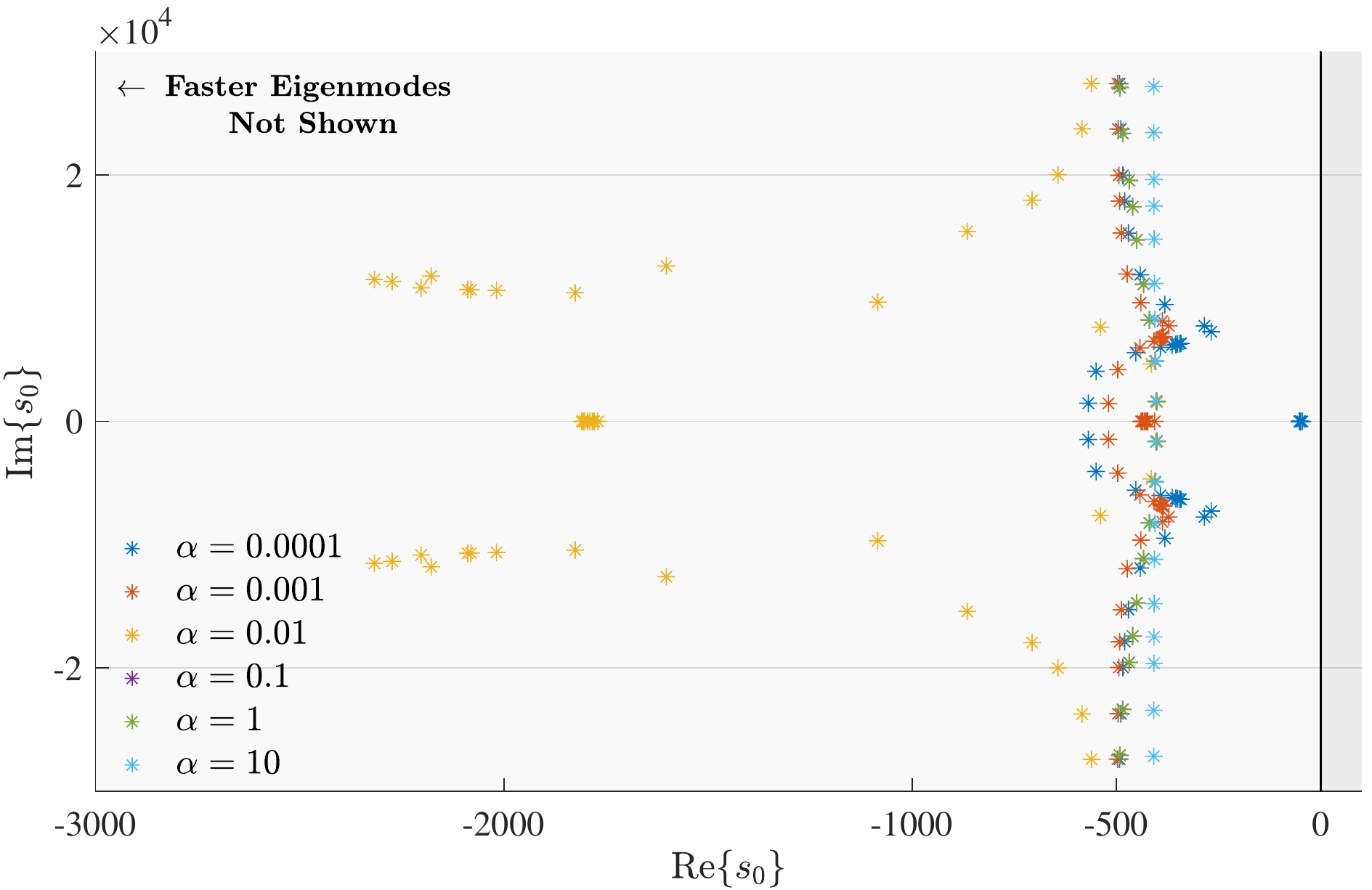}
    \caption{{Eigenmode locus of the 10-bus radial microgrid when all buck converters are outfitted with lead-lag controllers. The eigenmodes are numerically computed via (\ref{eq: s_vec}): as the homotopy parameter $\alpha$ is scaled up, the eigenmodes of the system move in the complex plane. Since no eigenmodes move into the RHP, stability is never lost for $\alpha \le 10$.}}
    \label{fig:RL_LL_Radial}
\end{figure}

Next, we confirm these results using Lemma \ref{lemma:eig_mode} by testing the system eigenmodes on a root locus plot as $\alpha$ is scaled. To do so, we build the parameterized system admittance matrix ${\bf Y}(s,\alpha)\in{\mathbb C}^{10\times10}$ and construct the $N^{\rm th}$ order polynomial ${\rm det}[{\bf Y}(s,\alpha)]=0$, where $N=54$. Finally, we employ the MATLAB root solver $\mathtt{roots}(\cdot)$ to build the complex vector 
\begin{align}\label{eq: s_vec}
{\bf s}(\alpha)=\{s\;|\;{\rm det}[{\bf Y}(s,\alpha)]=0\}.
\end{align}
As shown in Fig. \ref{fig:RL_LL_Radial}, no eigenmodes cross into the RHP for any value of $\alpha \le 10$. Therefore, the conclusion of Corollary \ref{corollary: phase_simp2} is correct: the network is stable for any gain value $G_{c\infty}$, even up to 10 times its nominal value. Of course, we have confirmed the prediction only for a single network configuration, but the results should generalize to an arbitrary network configuration. 

\begin{figure}
    \centering
    \includegraphics[width=1\linewidth]{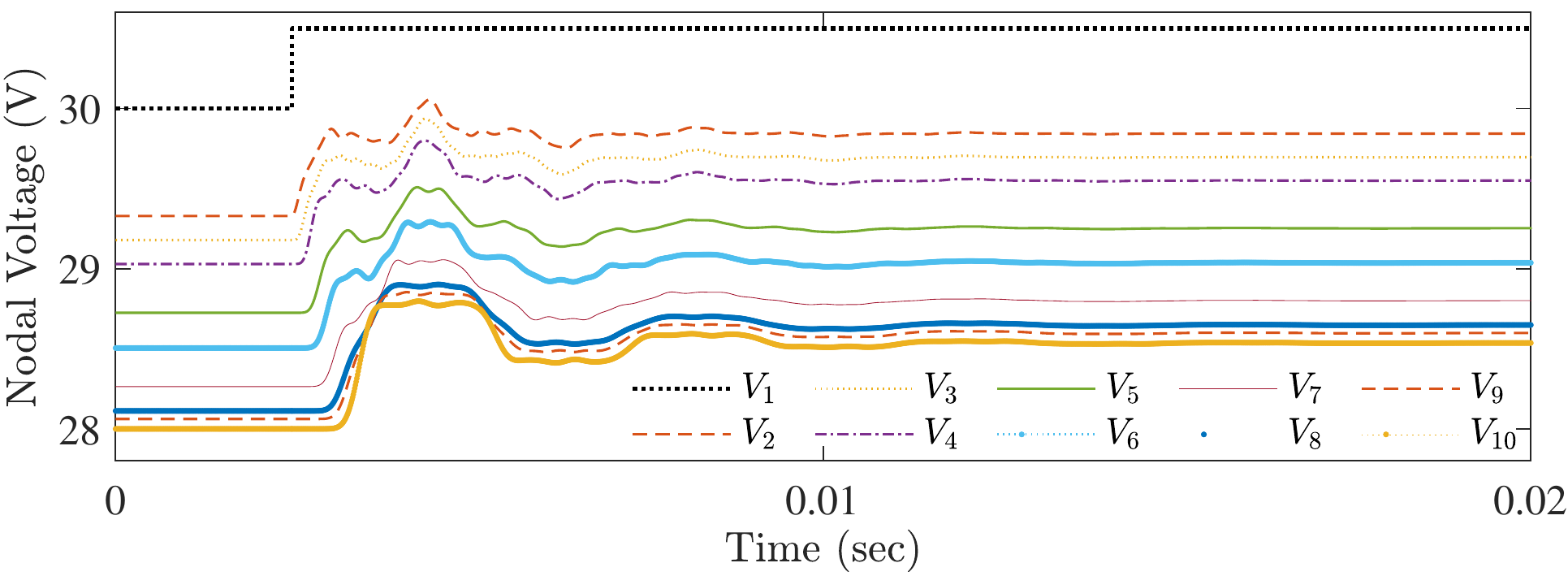}
    \caption{{Shown are the nodal voltage dynamic responses for a 0.5 V step change in supply voltage at source bus 1. Despite the large controller gain values associated with $\alpha=10$, system settling time is on the order of 10ms.}}\label{fig:TD_PlotsLL_Radial}
\end{figure}

{To further confirm this result, we ran a time domain simulation of the microgrid dynamics in response to a small perturbation. Accordingly, we set $\alpha=10$ and applied a 0.5 Volt step increase\footnote{{We assume this perturbation is sufficiently small such that the response is dominated by the linear component of the dynamics.}} to the voltage source set point at bus 1, which is a source bus in Fig. \ref{fig: MG_Power}. We then simulated the response in the time domain. The resulting dynamic responses, which are clearly stable, are shown in Fig.~\ref{fig:TD_PlotsLL_Radial}. In this simulation, we used average value models~\cite{erickson2007fundamentals} for all converter dynamics.}

\subsection{Stability Test{: 8-Bus Meshed Network}}
\begin{figure}
    \centering
    \includegraphics[width=1\linewidth]{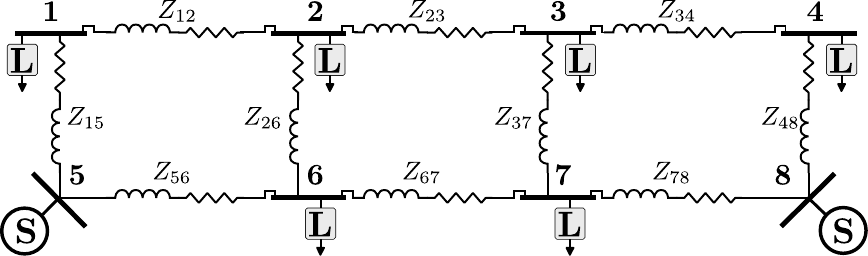}
    \caption{{Shown is a} meshed 8-bus microgrid with two stiff voltage sources and six loads. {Each load is interfaced through a converter.}}
    \label{fig:sim_setup}
\end{figure}

{We now preform a topology change: from the 10-bus radial network of Fig. \ref{fig: MG_Power} to the 8-bus meshed microgrid system depicted in Fig. \ref{fig:sim_setup}. Once again, all loads are equipped with buck converters, and we consider the stability of the network when the converters are under the influence of lead-lag compensators as in (\ref{eq: BCL_TF}). Stability assessment is also performed using an exact, centralized approach (i.e., root locus). Since the components in the network have not changed, the decentralized result derived via Corollary \ref{corollary: phase_simp2} from the previous subsection is still valid.} 

\begin{figure}
    \centering
    \includegraphics[width=1\linewidth]{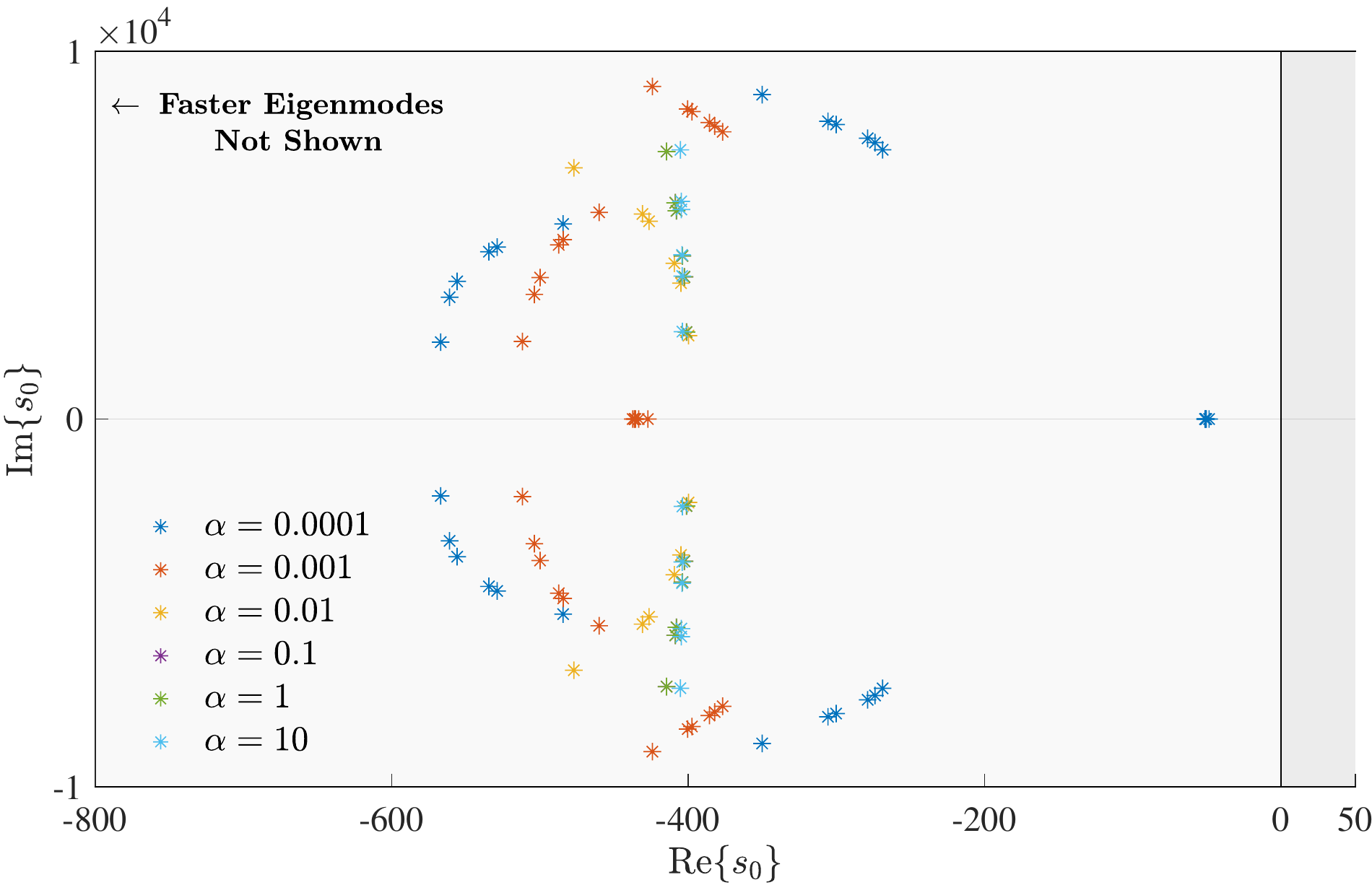}
    \caption{Eigenmode locus of the meshed 8-bus microgrid when all buck converters are outfitted with lead-lag controllers. {The eigenmodes are numerically computed via (\ref{eq: s_vec}). Since no eigenmodes move into the RHP,} stability is never lost for $\alpha \le 10$.}
    \label{fig:RL_LL}
\end{figure}

We test the stability of this meshed network via eigenmode locus by scaling $\alpha$. To do so, we build the parameterized system admittance matrix ${\bf Y}(s,\alpha)\in{\mathbb C}^{8\times8}$ and construct the $N^{\rm th}$ order polynomial ${\rm det}[{\bf Y}(s,\alpha)]=0$, where $N=40$, and apply (\ref{eq: s_vec}). As shown in Fig. \ref{fig:RL_LL}, no eigenmodes cross into the RHP for any value of $\alpha \le 10$. Therefore, the conclusion of Corollary \ref{corollary: phase_simp2} is once again correct, even under a significant topology change: the network is stable for any gain value $G_{c\infty}$, even up to 10 times its nominal value.

\begin{figure}
    \centering
    \includegraphics[width=1\linewidth]{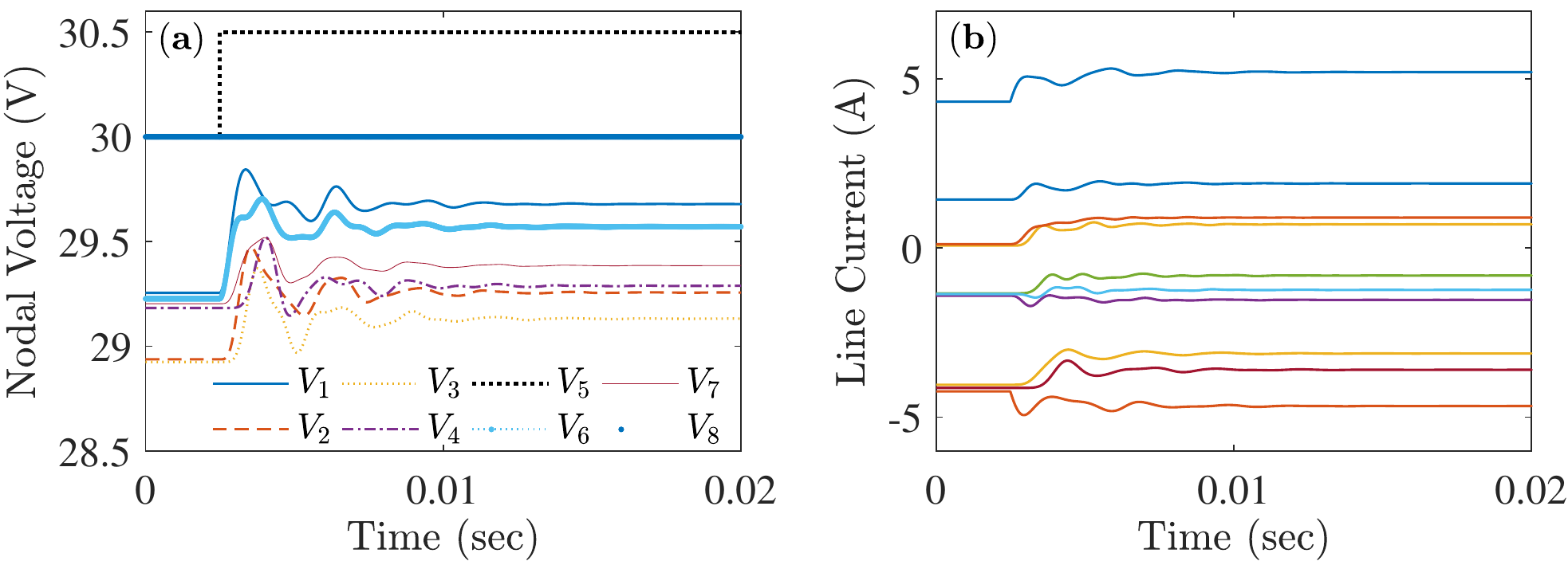}
    \caption{Shown are the nodal voltage (panel ($\bf a$)) and line current (panel ($\bf b$)) dynamic responses for a 0.5V step change in supply voltage at source bus 5. Despite the large controller gain values associated with $\alpha=10$, system settling time is on the order of 10ms.}    \label{fig:TD_PlotsLL}
\end{figure}

To further confirm this result, {we ran a time domain simulation of the microgrid dynamics in response to a small perturbation.} Accordingly, we set $\alpha=10$ and applied a 0.5 Volt step increase to the voltage source set point at bus 5, {which is a source bus in Fig. \ref{fig:sim_setup}. We then simulated the response in the time domain.} The resulting dynamic responses, {which are clearly stable}, are shown in Fig.~\ref{fig:TD_PlotsLL}.

{\subsection{Testing Network Stability: Buck \& Boost Converter Loads}
We now consider the stability of a microgrid system containing loads which are interfaced through both buck and boost converters. The small-signal model associated with the boost converter, the associated controller tuning, and its homotopic parameterization are all given in Appendix \ref{Boost_Converter}. In Fig. \ref{fig:BuckBoost}, we plot the sector curves associated with the line, buck, and boost converters for various parameterization values. Since there exists continuous overlap between all three of these devices, the elements are able to collectively satisfy Corollary \ref{corollary: phase_simp2} up to the parameterization value $\alpha=1.25$. Thus, any configuration of these devices is guaranteed to be plug-and-play stable.}

\begin{figure}
    \centering
    \includegraphics[width=1\linewidth]{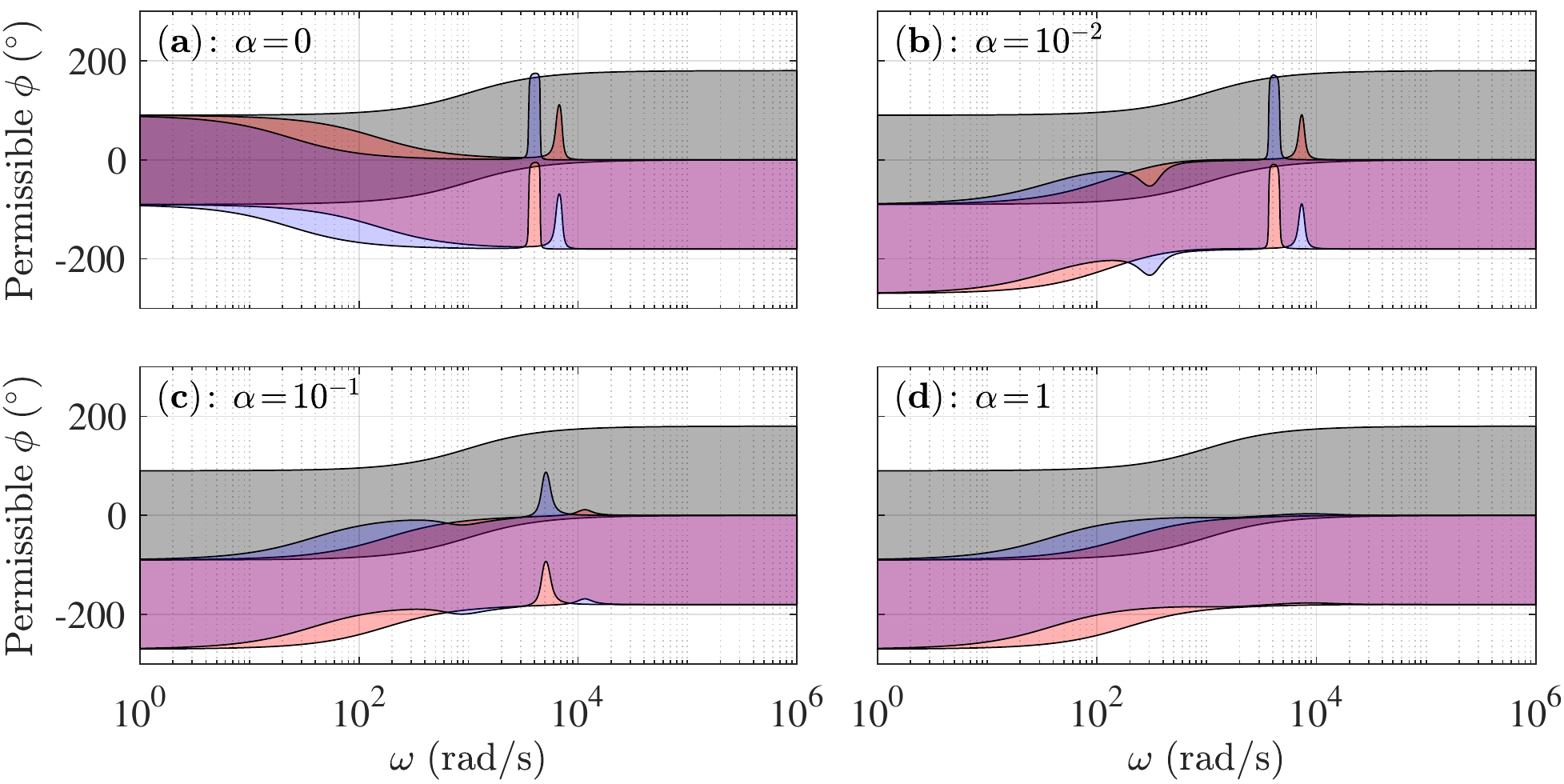}
    \caption{{Line, Buck (Lead/Lag Compensator) Converter, and Boost Converter Sector Plots. Plotted are the values that $\phi(\omega,\alpha)$ can take, according to (\ref{eq: phase_cond2}), for a continuum of frequency and gain values. The grey sector corresponds to the line phase function (\ref{eq: line_phase}), the red sector corresponds to the buck converter load phase function (\ref{eq: buck_phase}), and the blue sector corresponds to the boost converter load phase function. Note: the overlap between the red and blue sectors appears purple. For all considered values of $\alpha$, there exists continuous overlap between all three sectors.}}
    \label{fig:BuckBoost}
\end{figure}

\begin{figure}
    \centering
    \includegraphics[width=1\linewidth]{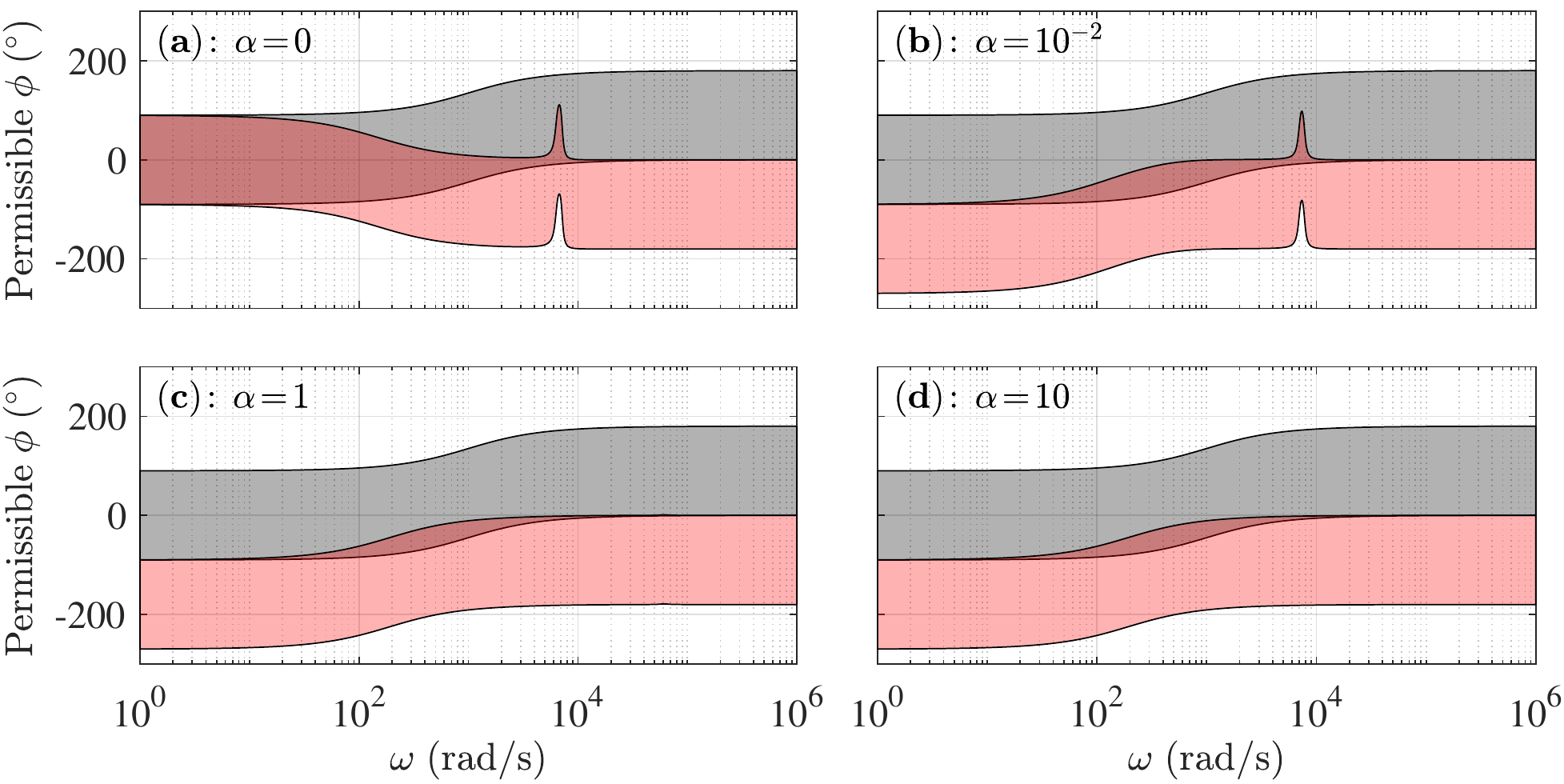}
    \caption{{Line and Buck Converter (Delayed Lead/Lag Compensator) Sector Plots. Plotted are the values that $\phi(\omega,\alpha)$ can take, according to (\ref{eq: phase_cond2}), for a continuum of frequency and gain values. The grey sector corresponds to the line phase function (\ref{eq: line_phase}), while the red sector corresponds to the load phase function (\ref{eq: buck_phase}), where $G_c(s)$ is a lead-lag controller. Despite the time delay applied to the loop gain via (\ref{eq: delay}), Corollary \ref{corollary: phase_simp2} can still be satisfied for this level of delay, so plug-and-play stability is preserved.}}
    \label{fig:BuckLL_Curves_Delay}
\end{figure}

{\subsection{The Effects of Time Delay}
The effects of time delay~\cite{Baghaee:2017} can be incorporated into the application of Corollary \ref{corollary: phase_simp2} without difficulty. For example, the effect of time delay in the power electronic converters can be considered by adding an exponential delay term directly into the loop gain equation. That is,
\begin{align}\label{eq: delay}
    T=\left(G_{vd}(s)H(s)G_{c}(s)/V_{m}\right)e^{-s\tau_{d}},
\end{align}
where $\tau_d$ is the delay time. For a 100 kHz switching frequency, $\tau_d=10^{-5}$ is a reasonable delay estimate. Next, we can numerically recalculate the phase of the converter via (\ref{eq: buck_phase}). To showcase this capability, we repeat the analysis associated with Fig. \ref{fig:BuckLL_Curves}. That is, we analyze a line and a buck converter outfitted with a delayed lead/lag compensator. The results are shown in Fig. \ref{fig:BuckLL_Curves_Delay}. Based on these results, we can conclude that the associated network remains plug-and-play stable for this value of $\tau_d$.}

{\subsection{Comparisons to Other Methods}\label{subsec: comparisons}
In this subsection, we qualitatively compare our method to other leading stability assessment techniques for DC microgrid systems, namely, the Middlebrook Criterion, the Gain Margin Phase Margin (GMPM) Criterion, and the Opposing Argument Criterion\cite{riccobono2012comprehensive}. These criteria are applied at the interconnection of two systems with their effective impedances $Z_0(s)$ and $Z_i(s)$;  we will apply them to assess the stability of connecting a buck converter to bus No.$1$ in the grid of Fig.~\ref{fig:sim_setup}, assuming converters on all the other buses are already in place.}

{We note that the results of the application of these methods (Middlebrook, GMPM, and Opposing Argument) should not be directly quantitatively compared to the results of our method. The reason is that these methods require the full knowledge of the network topology to be applied, thus, they certify ``compatibility" of the components for a specific interconnection. Our proposed method, on the contrary, provides a ``network topology invariant" certificate, thus ensuring compatibility of the chosen components under arbitrary interconnections.}         

\begin{figure}
    \centering
    \includegraphics[width=0.85\linewidth]{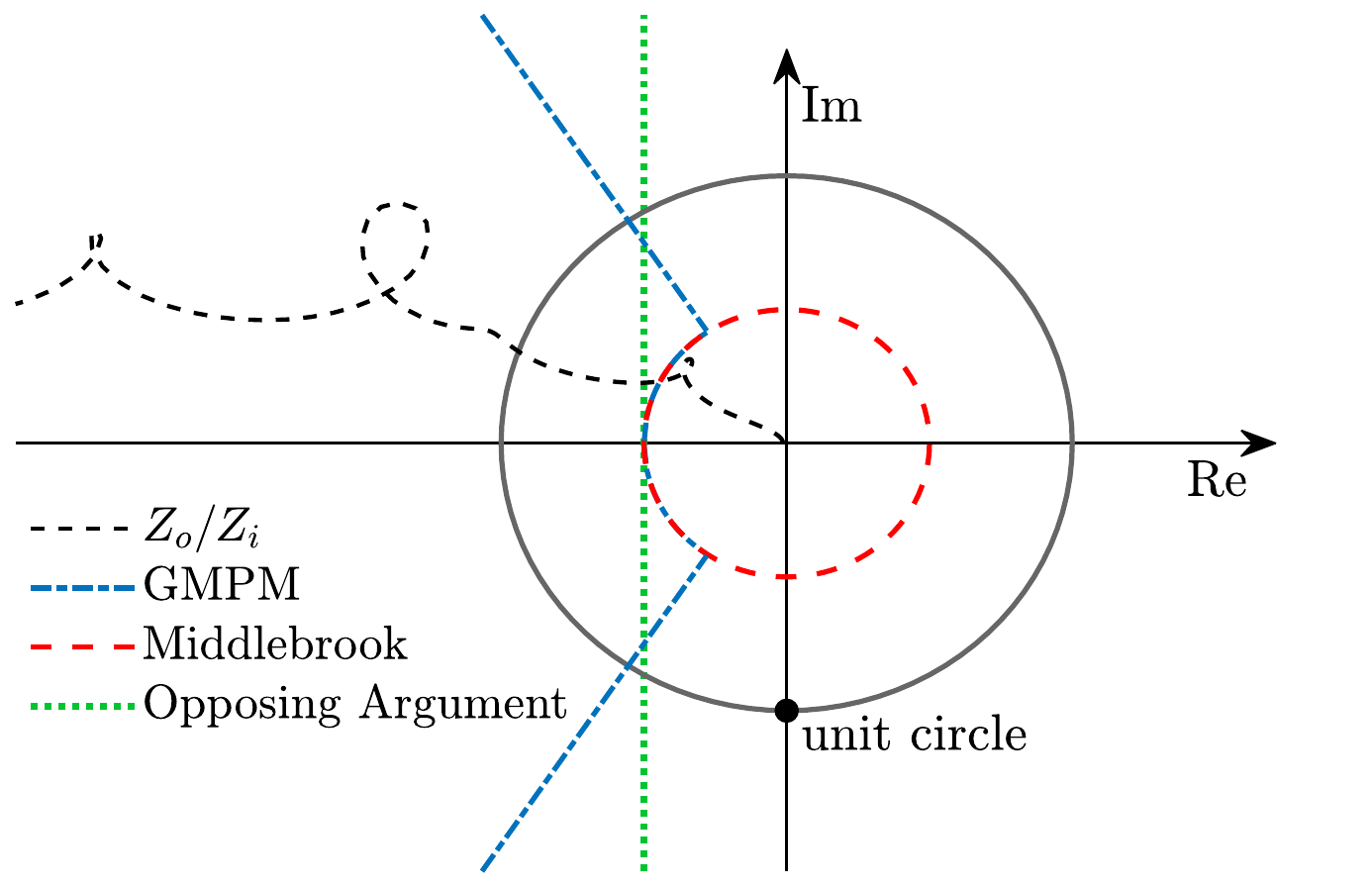}
    \caption{{Stability assessment via Middlebrook, GMPM, and Opposing Argument criteria for connecting a buck converter to bus No.$1$ of the microgrid of Fig.~\ref{fig:sim_setup}. The dashed line represents the ratio of the grid and converter impedance, while the red, blue, and green lines show the boundaries set by the Middlebrook, GMPM, and Opposing Argument Criterion, respectively. None of the criterion are satisfied.}}\label{fig:middle_crit}
\end{figure}

{In order to apply the listed stability criteria, we first need to calculate the effective admittance of the grid (including converters) as seen from bus No.$1$. Accordingly, we determine this effective admittance by applying a ``test" current at bus No.1; then, equation \eqref{eq: 0=Yv} reads: 
\begin{align}
\left[\begin{array}{c}
i_{1}(s)\\
\hline {\bf 0}
\end{array}\right]=\left[\begin{array}{c|c}
Y_{1}(s) & \bm{Y}_{2}(s)\\
\hline \bm{Y}_{3}(s) & \bm{Y}_{4}(s)
\end{array}\right]\left[\begin{array}{c}
v_{1}(s)\\
\hline \bm{v}_{n}(s)
\end{array}\right].
\end{align}
Here $i_1$ and $v_1$ are the current injection and voltage at bus No.$1$ (scalars), while $\bm{v}_{n}(s)$ it the vector of voltages on all the other buses. We may solve for the effective output admittance ${Y_{o}(s)}$ as seen by the injection current~\cite{machowski2011power}:
\begin{align}
i_{1}(s)=\underbrace{\left(\bm{Y}_{1}(s)-\bm{Y}_{2}(s)\bm{Y}_{4}(s)^{-1}\bm{Y}_{3}(s)\right)}_{Y_{o}(s)}v_{1}(s).
\end{align}
Fig.~\ref{fig:middle_crit} provides the result of the application of the Middlebrook Criterion, the GMPM Criterion, and the Opposing Argument Criterion. We note, that none of them are satisfied, which means that for this particular system, stability cannot be satisfied by applying them, although the system is in fact stable. We further note, however, that the methods applied provide stability certificates with certain conservativeness, so the fact that they are not satisfied does not mean that the system should be unstable. However, it is clear that our method performs well and certifies stability of the system, even when conventional methods fail to do so.}

{\section{Controller Adjustments for Stability Certification}\label{sec:control_adjust}
In the present section, we will consider how one can use the developed method to adjust component controllers to achieve plug-and-play system stability. For this, we specifically chose a controller setting such that the condition of Corollary \ref{corollary: phase_simp2} is not satisfied initially. Then, we show how one can change the controller gains so that this condition is satisfied, thus certifying stability. We consider two different configurations: PI controlled buck converters, and lead-lag buck converters as in the previous section, but in combination with much more inductive power lines.}

\subsubsection{PI Controller}
\begin{figure}
    \centering
    \includegraphics[width=1\linewidth]{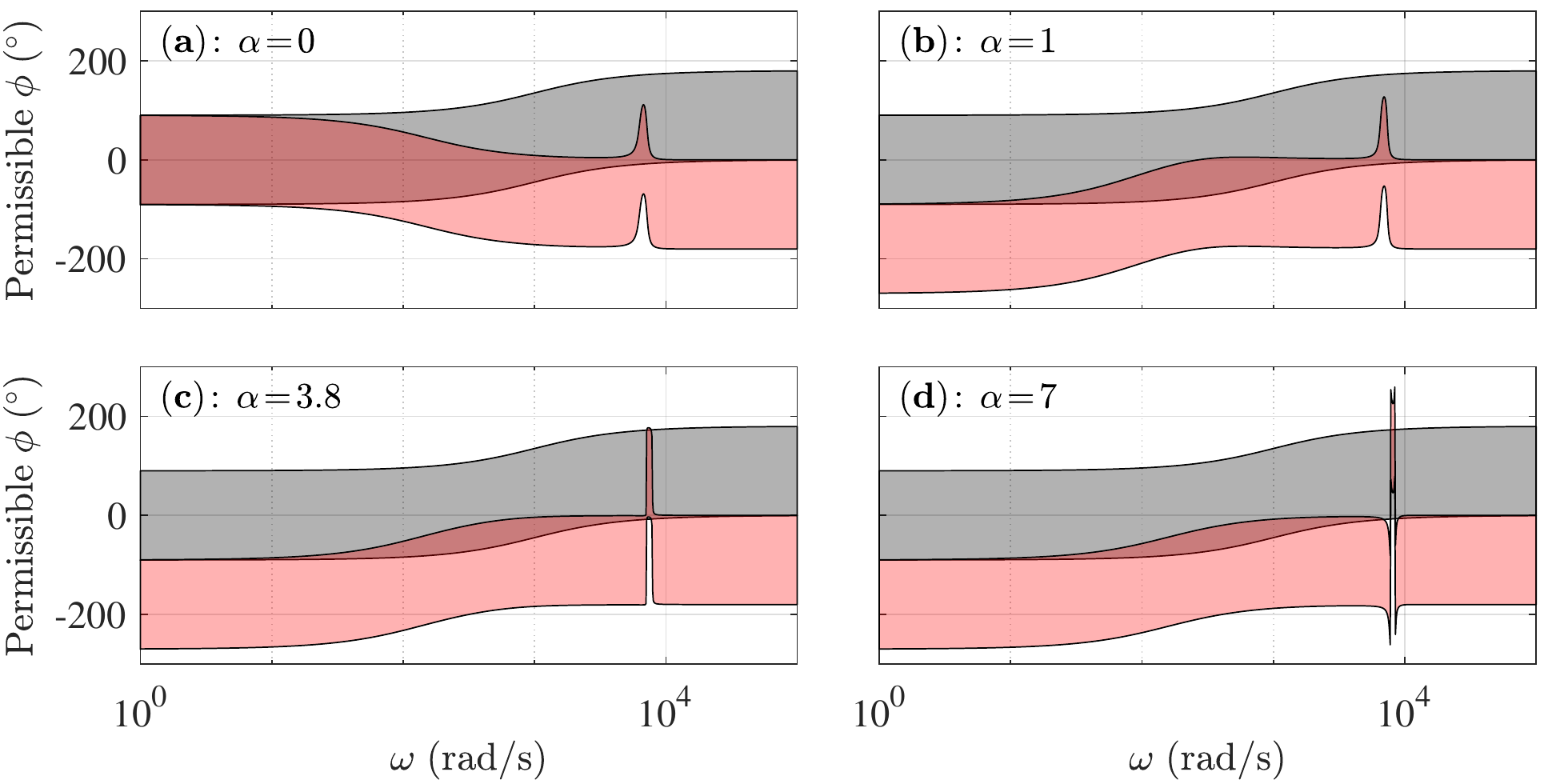}
    \caption{{Line and Buck Converter (PI Compensator) Sector Plots.} Plotted are the values that $\phi(\omega,\alpha)$ can take, according to (\ref{eq: phase_cond2}), for a continuum of frequency and gain values. The grey sector corresponds to the line phase function (\ref{eq: line_phase}), while the red sector corresponds to the load phase function (\ref{eq: buck_phase}), where $G_c(s)$ is a PI controller.}
    \label{fig:BuckPI_Curves}
\end{figure}

\begin{figure}
    \centering
    \includegraphics[width=1\linewidth]{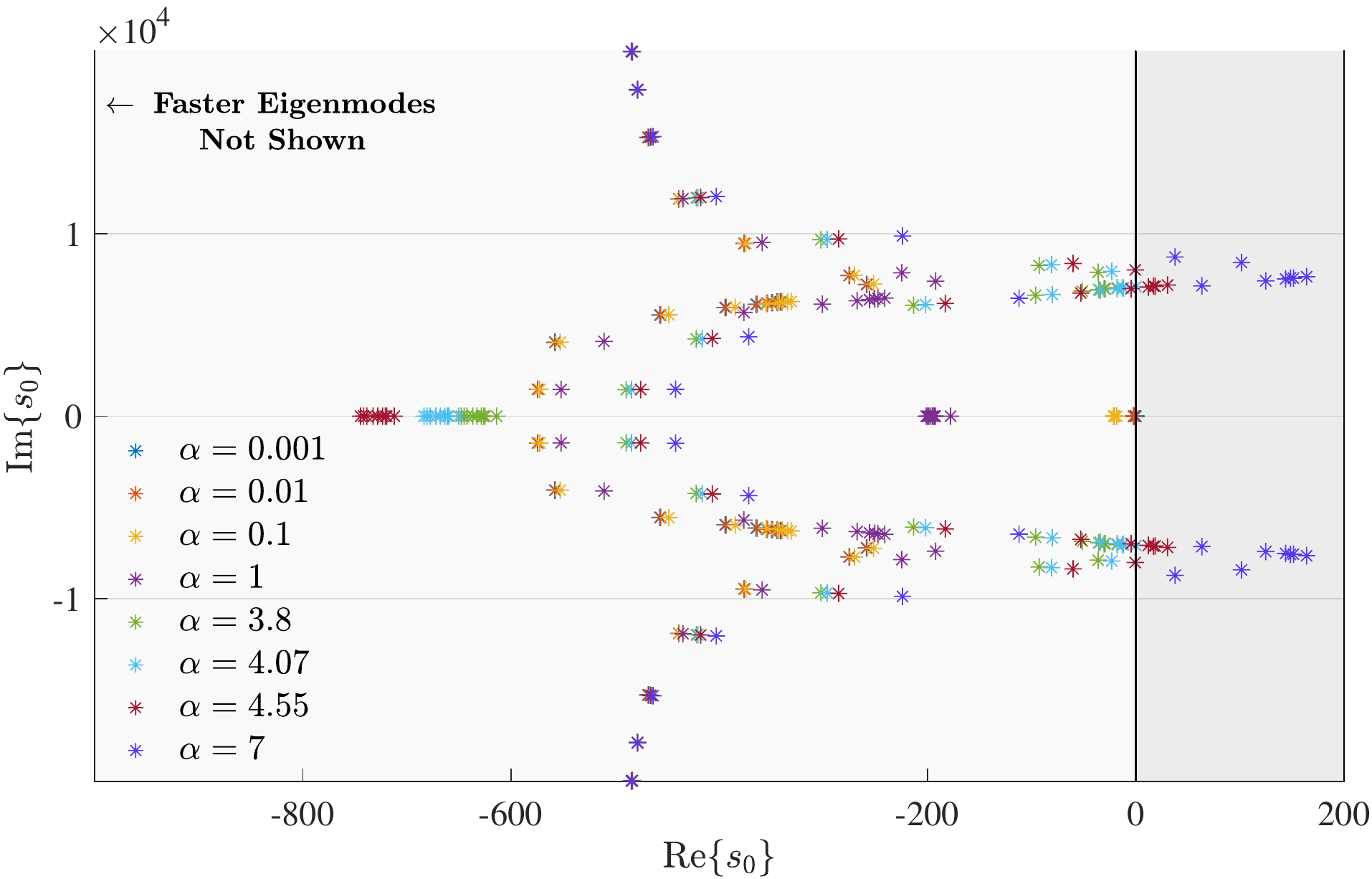}
    \caption{{Eigenmode locus of the 10-bus radial microgrid when all buck converters are outfitted with PI controllers. The eigenmodes are numerically computed via (\ref{eq: s_vec}). Since a pair of eigenmodes cross into the RHP when $\alpha=4.07$, system stability is lost for $\alpha>4.07$.}}
    \label{fig:RL_PI_Radial}
\end{figure}

In this test, we consider the $10$-bus radial network with buck converters with PI controllers:
\begin{align}\label{eq: PI_c}
G_{c}(s)=G_{i}\left(1+\tfrac{\omega_{i}}{s}\right).
\end{align}
Setting $\omega_i=2\pi 500$ and $G_i=0.015$, the phase and gain margins of $1+T(s)$ were driven to $\phi_m=93.6^{\circ}$ and $g_m=12.3$ dB. The resulting phase function analysis results are shown in Fig. \ref{fig:BuckPI_Curves}, where the phase functions are plotted from $\alpha=0$ up to $\alpha=7$ (i.e., scaling the gain $G_{c\infty}$ to 7 times its nominal value). As can be seen, there is a break in continuous overlap between the sectors at $\alpha=3.8$, corresponding to a \textit{maximum allowable controller gain} of $3.8\times0.015=0.057$. The root locus plot is shown in Fig. \ref{fig:RL_PI_Radial}. The network is clearly still stable for $\alpha=3.8$, but becomes marginally stable for $\alpha=4.07$. Stability is clearly lost for larger value of $\alpha$, as shown by the eigenmodes of $\alpha=7$, for example. Therefore, Corollary \ref{corollary: phase_simp2} is conservative by a margin of $100\times(1 - 3.8/4.07)\approx 6.6\%$ \textit{for this particular network configuration}. In this context, the results of Fig. \ref{fig:BuckPI_Curves} can be interpreted alternatively: there exists some network configuration for which $\alpha=3.8$ is the maximum allowable gain factor.

\begin{figure}
    \centering
    \includegraphics[width=1\linewidth]{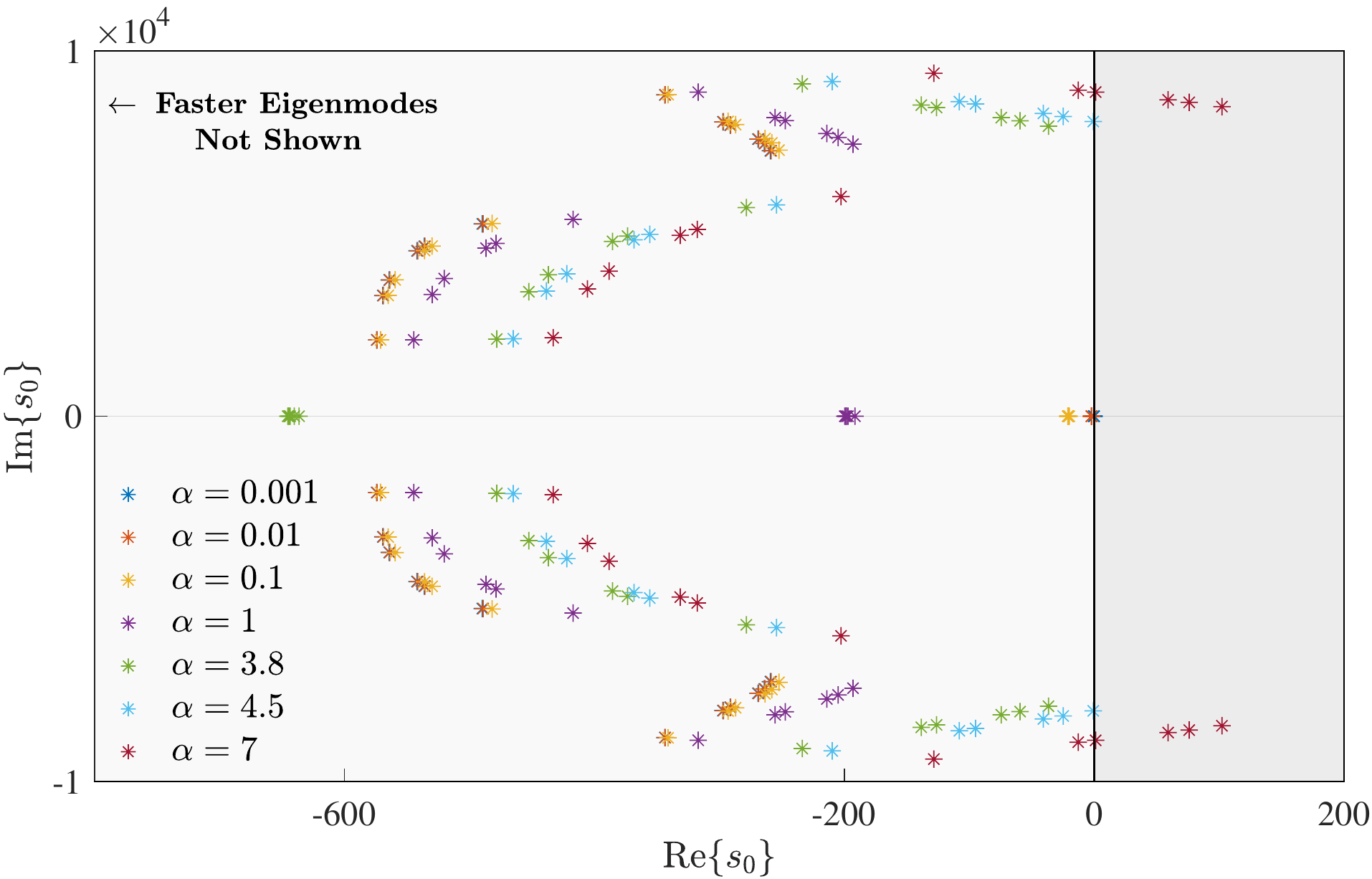}
    \caption{Eigenmode locus of the meshed 8-bus microgrid when all buck converters are outfitted with PI controllers. {The eigenmodes are numerically computed via (\ref{eq: s_vec}). Since a pair of eigenmodes cross into the RHP when $\alpha=4.5$,} system stability is lost for $\alpha>4.5$.}
    \label{fig:RL_PI}
\end{figure}

Let us now test the same buck converters with PI controllers (\ref{eq: PI_c}), but under the meshed $8$-bus network of Fig.~\ref{fig:sim_setup}. The corresponding eigenmode locus plot is shown in Fig. \ref{fig:RL_PI}. The network is clearly still stable for $\alpha=3.8$, but becomes marginally stable for $\alpha=4.5$. Stability is clearly lost for larger value of $\alpha$, as shown by the eigenmodes of $\alpha=7$, for example. Therefore, Corollary \ref{corollary: phase_simp2} is conservative by a margin of $100\times(1 - 3.8/4.7)\approx 20\%$ \textit{for this particular network configuration}. {We notice, therefore, that Corollary \ref{corollary: phase_simp2} was less conservative in the radial case than it was in the meshed case.}

{\subsubsection{Lead-Lag controller and inductive lines}
Here we consider buck converters with lead-lag voltage compensators (as in the previous section) in a network with very inductive lines. It is generally known that inductive lines promote instability, since the electromagnetic process timescale becomes larger, and converter controllers become too fast to operate stably. For the purpose of illustrating of how controllers can be adjusted to stabilize the system, we choose the electromagnetic time to be $\tau=6$ms.}

\begin{figure}
    \centering
    \includegraphics[width=1\linewidth]{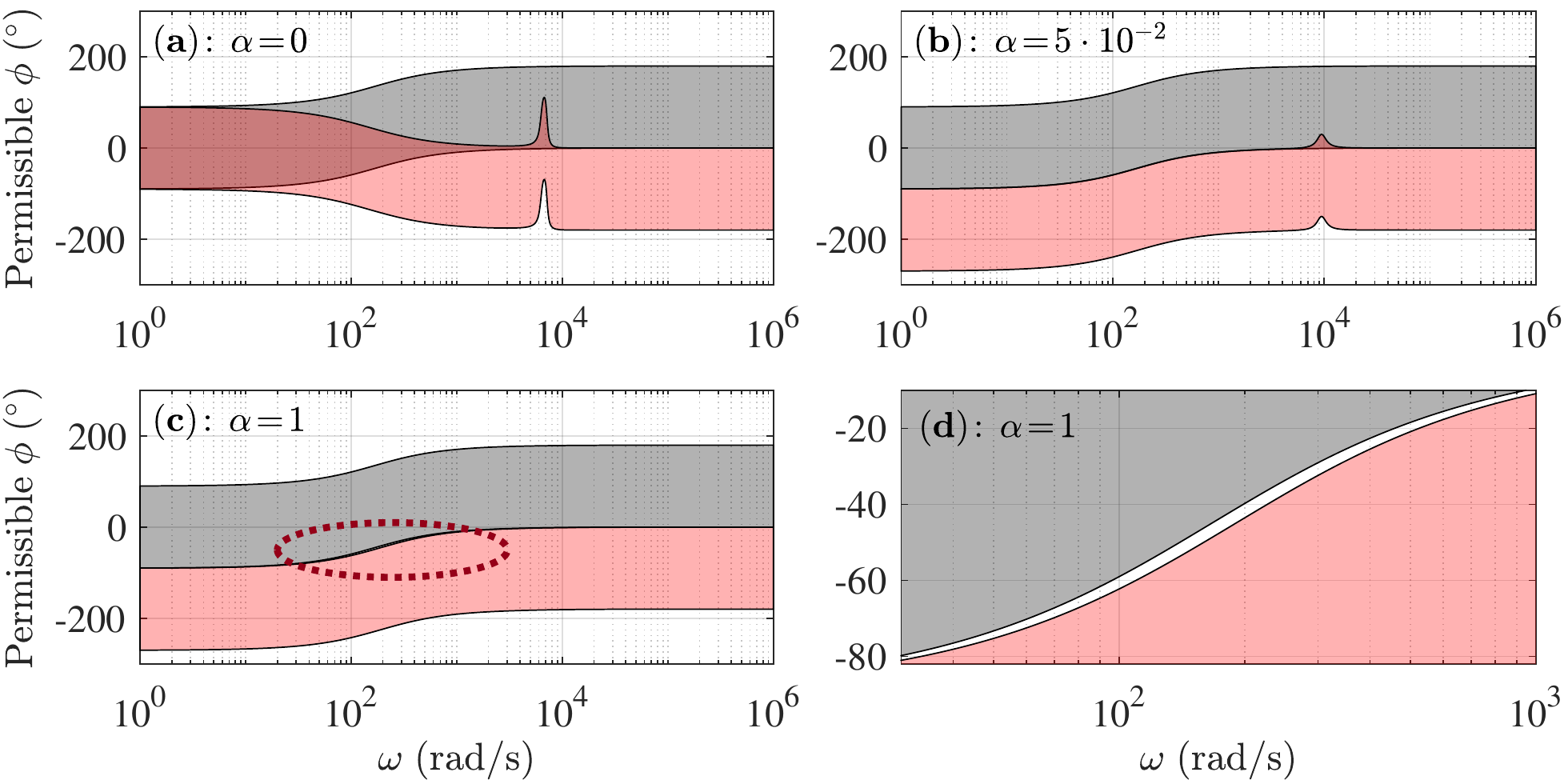}
    \caption{{Line and Buck Converter (Lead/Lag Compensator) Sector Plots.} Plotted are the values that $\phi(\omega,\alpha)$ can take, according to (\ref{eq: phase_cond2}), for a continuum of frequency and gain values. It can be infered, that for $\alpha=1$ there exist a non-overlap region. The maximum value of $\alpha$ that still guarantees overlap is around $\alpha \approx 0.05$, which provides the factor for controller gain adjustment.}\label{fig:tau6ms}
\end{figure}

{Fig.~\ref{fig:tau6ms} shows that the phase condition of Corollary $2$ is not satisfied for the chosen controller configuration and line parameters. One can find that the maximum value of the scaling parameter $\alpha$, such that the condition of Corollary \ref{corollary: phase_simp2} is everywhere satisfied, is about $\alpha \approx 0.05$. This implies that if the lead-lag controller gain is scaled down by this factor, the system becomes stable under arbitrary interconnections. This is equivalent to reducing the buck converter cross-over frequency from $\omega_c \approx 5$ kHz to $\omega_c \approx 300$ Hz, which is logical, since more inductive lines require slower controllers on converters.  }

\section{{Validation Through Real-Time Simulations}}\label{sec: RTS}

{In this section, we provide the validation of our results using switch-level models of power converters. We first present the results of simulations using Matlab/Simulink software, and then the real-time simulations using OPAL-RT system.} 

\begin{figure}
    \centering
    \includegraphics[width=1\linewidth]{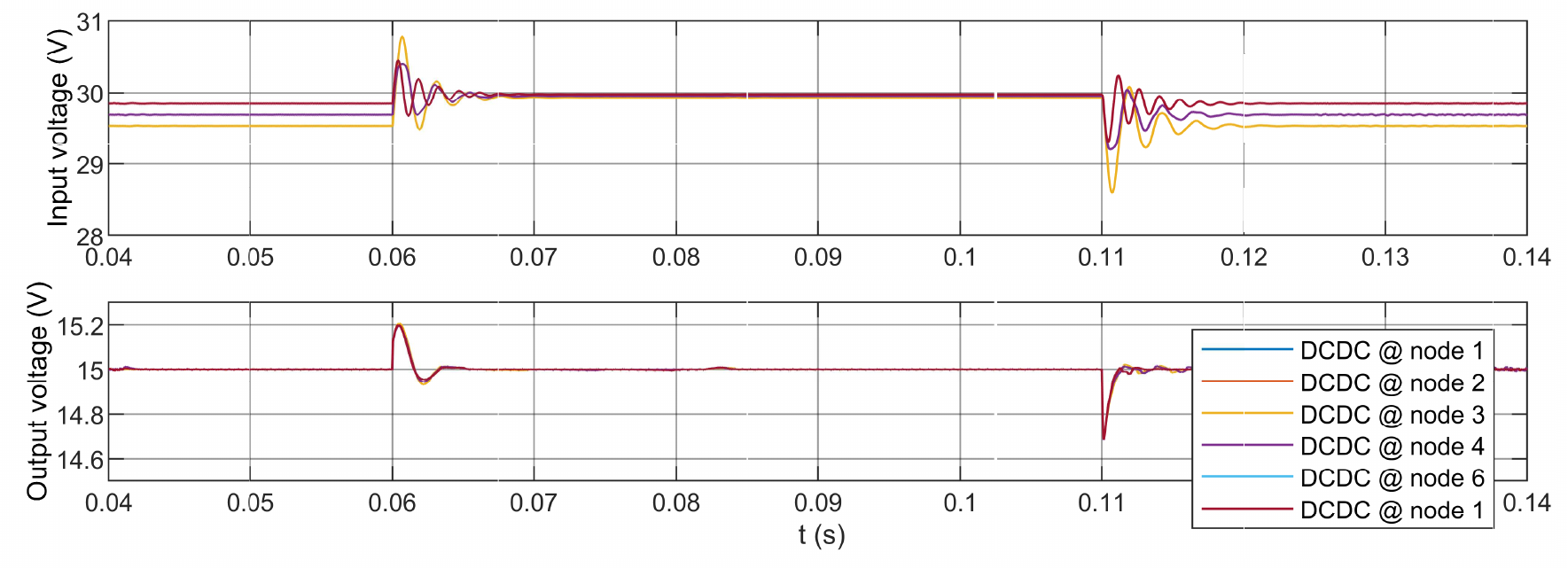}
    \caption{{Simulations of the network of Fig. \ref{fig:sim_setup} using Matlab/Simulink, switch-level converter models. Time-constants of the network lines: $\tau=1$ ms. }} \label{fig:simulink1}
\end{figure}

\begin{figure}
    \centering
    \includegraphics[width=1\linewidth]{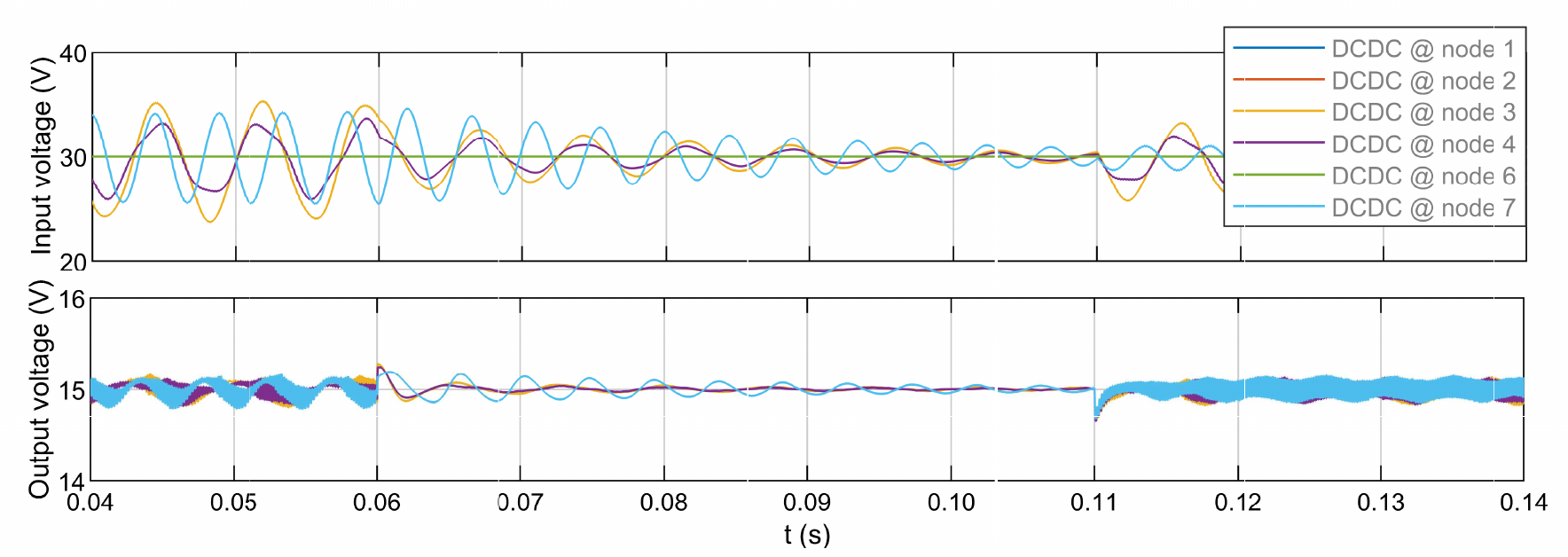}
    \caption{{Simulations of the network of Fig. \ref{fig:sim_setup} using Matlab/Simulink, switch-level converter models.  Time-constants of the network lines: $\tau=10$ ms.}} \label{fig:simulink10}
\end{figure}

\begin{figure}
    \centering
    \includegraphics[width=1\linewidth]{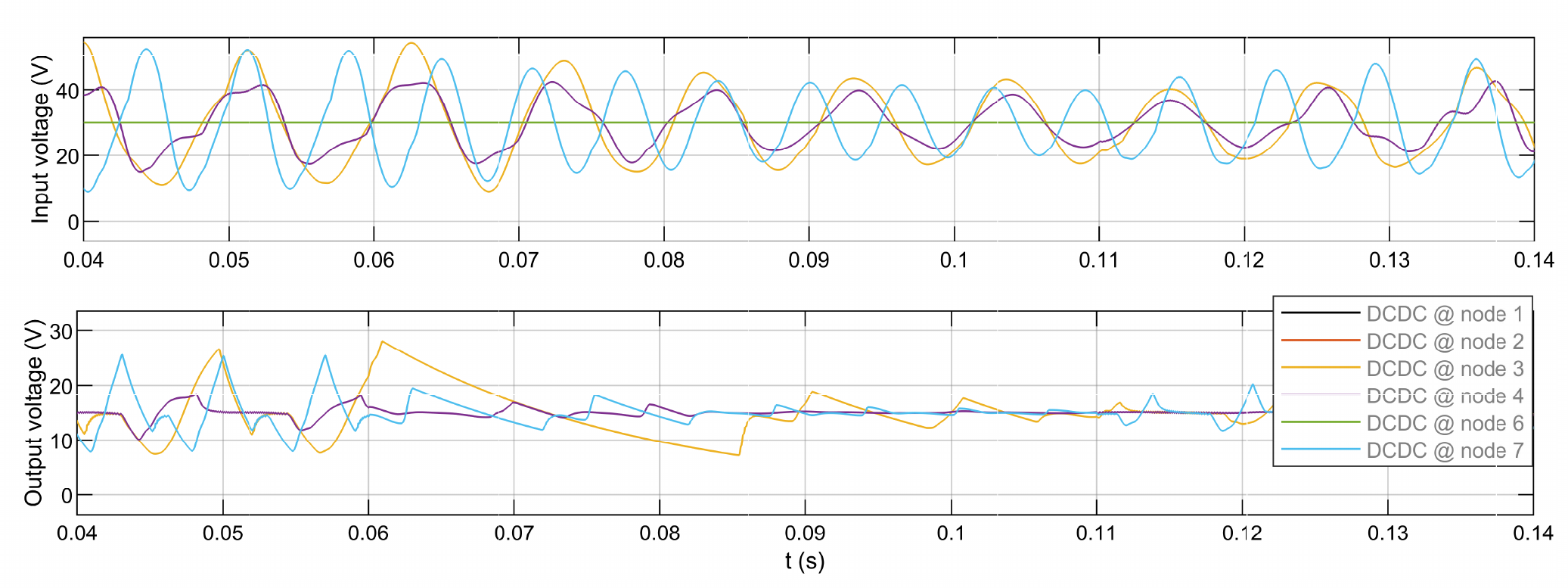}
    \caption{{Simulations of the network of Fig. \ref{fig:sim_setup} using Matlab/Simulink, switch-level converter models.Time-constants of the network lines: $\tau=20$ ms. }} \label{fig:simulink20}
\end{figure}

{Figures \ref{fig:simulink1}, \ref{fig:simulink10}, and \ref{fig:simulink20} present the results of dynamic simulations for the meshed system of Fig. \ref{fig:sim_setup} using Matlab/Simulink with switch-level models of buck converters, assuming different time-constants of the network lines: $\tau=1$ ms, $\tau=10$ ms, and $\tau=20$ ms, respectively. Converter parameters are the same in every case, and they are given in Table \ref{tab:sim_param_CC} (column, labeled ``NUM"). For all the three figures, a step load change is applied at the time instant $t=0.06$ a, and at the time $t=1.1$ s, the load is returned back. It is seen that for the line time constant $\tau=1$ms the systems is stable, as is predicted by our method. For the lines with the time-constant $\tau=10$ ms our method already does not guarantee stability -- it certifies stability only for line time-constants less than $\sim6$ ms. Figure \ref{fig:simulink10} shows that in this case the system is still formally stable, but one can already see rather large transients. Again, we note, that when our method can not guarantee stability, it means, that there exist \emph{some configuration} when the system is unstable. In this particular case, the network of the Fig. \ref{fig:sim_setup} is not the worst possible case, so it remains marginally stable. Next, Figure \ref{fig:simulink20} shows the case for the line time-constant $\tau=20$ ms. In this case, the system is unstable and substantial oscillations of both converter input and output voltage are present all the time.}    

\begin{figure}
    \centering
    \includegraphics[bb=40bp 0bp 495bp 207bp,clip,width=1\linewidth]{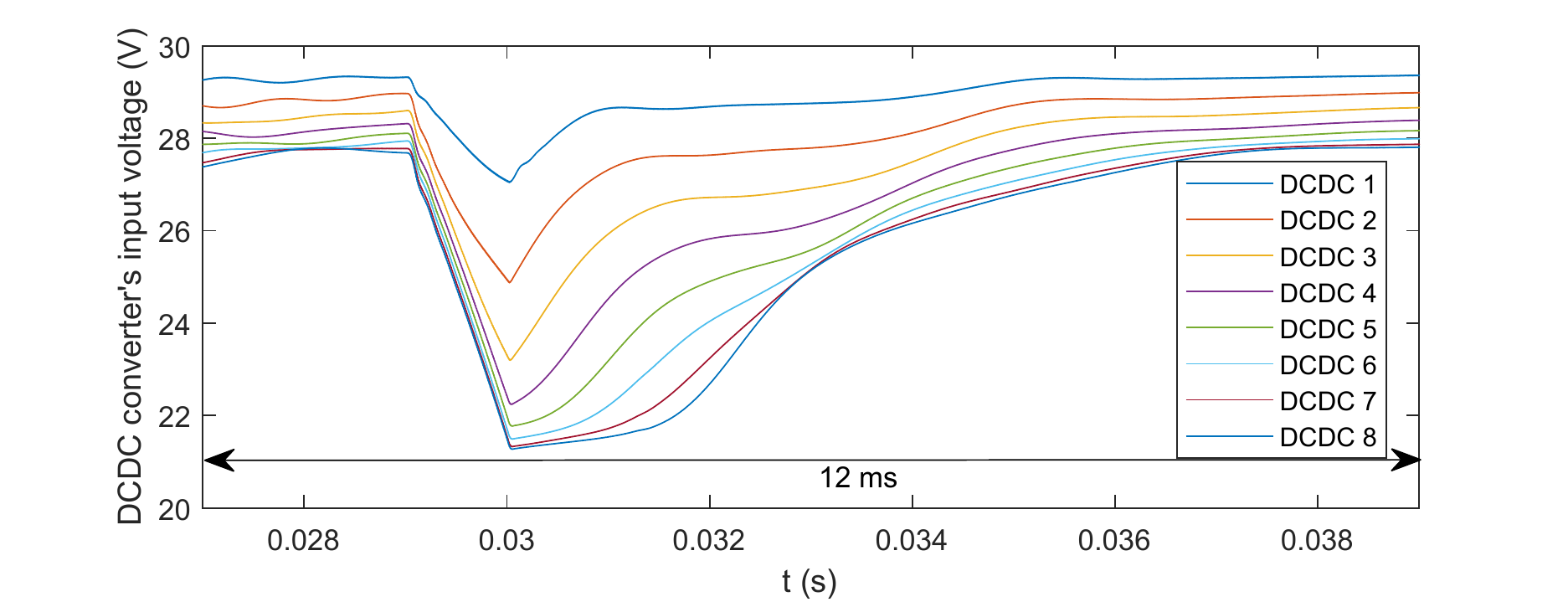}
   {(\textbf{a})}
    \includegraphics[bb=55bp 0bp 615bp 235bp,clip,width=1\linewidth]{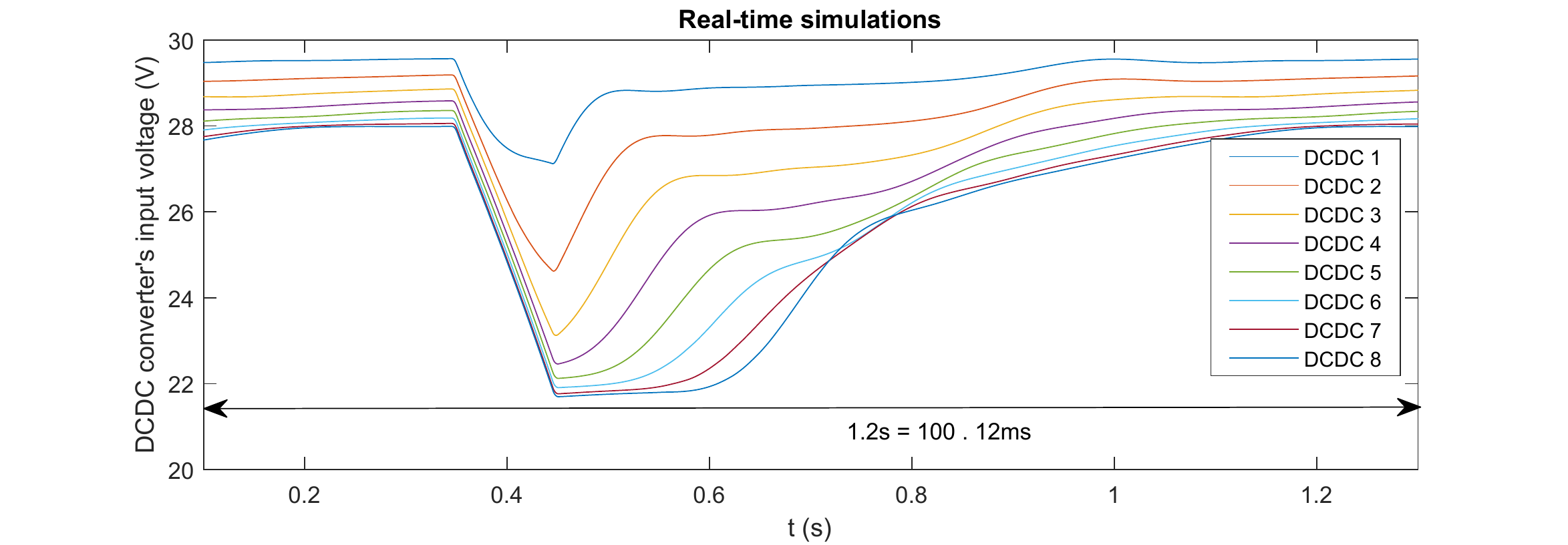}
    {(\textbf{b})}
    \caption{{Comparison of dynamic simulations over the radial network with $8$ buck converters: (\textbf{a}) simulation at $100$ kHz switching frequency, (\textbf{b}) scaled-down real-time simulation at $1$ kHz switching frequency.}} \label{fig:simvsrt}
\end{figure}

{Next, we move to perform validation via real-time simulations, using the OPAL-RT 5600 systems. Due to a time-step limitation of our OPAL-RT system that limits the switching frequency, we have scaled down the system time-scales by a factor of $k_f=100$. The scale down procedure consisted of using the switching frequency of $1$ kHz while keeping the same ripple in the converter’s inductor current, such that
\begin{align}\label{eq: scaledwn}
L \tfrac{\Delta i_L}{D T_S} = L' \tfrac{\Delta i_L}{D T_S'},
\end{align}
where $\Delta i$ is the current ripple which is the same as in the original setup. Parameters $L$, $T_s$ are the original inductance and switching period, while $L’$ and $T_s’$ are the scaled down inductance and switching period, and $D$ is the duty cycle. $T_s’$ is defined as $T_s’=k_f \cdot Ts$, so using (\ref{eq: scaledwn}), $L’=L\cdot k_f$.
In addition, the frequency response of the converter must be similarly transformed down by the same factor ($k_f$ times), such that the resonant frequency is
\begin{align}\label{eq: Cscaledwn}
\tfrac{1}{\sqrt{LC}} = k_f \tfrac{1}{\sqrt{L’C’}}. \end{align}
Thus, the capacitance must be $C’=k_f C$. Finally, all zeros and poles of the controller must be scaled down by $k_f$ too. In order to validate this scaling approach, a simulation in Matlab/Simulink at the $100$kHz switching frequency was compared with a real-time simulation using OPAL-RT at the $1$ kHz switching frequency. Fig. \ref{fig:simvsrt} presents the response of $8$ buck converters in a radial network (see Fig. 1) configuration; the response is generated by a step change in the load of all of the converters simultaneously (from half loaded to fully loaded). Panel (\textbf{a}) of Fig. \ref{fig:simvsrt} shows that the transient response lasts $12$ ms in the Matlab/Simulink simulation case. Equivalent results are measured from the real-time simulator, but they last 1.2 s, according to our scaling choice (see Fig. \ref{fig:simvsrt}, panel (\textbf{b})). The parameters associated with the converters and the network are presented in Table \ref{tab:sim_param_CC}, under the column labeled ``RTS".}

\begin{figure}
    \centering
    \includegraphics[bb=55bp 0bp 620bp 209bp,clip,width=1\linewidth]{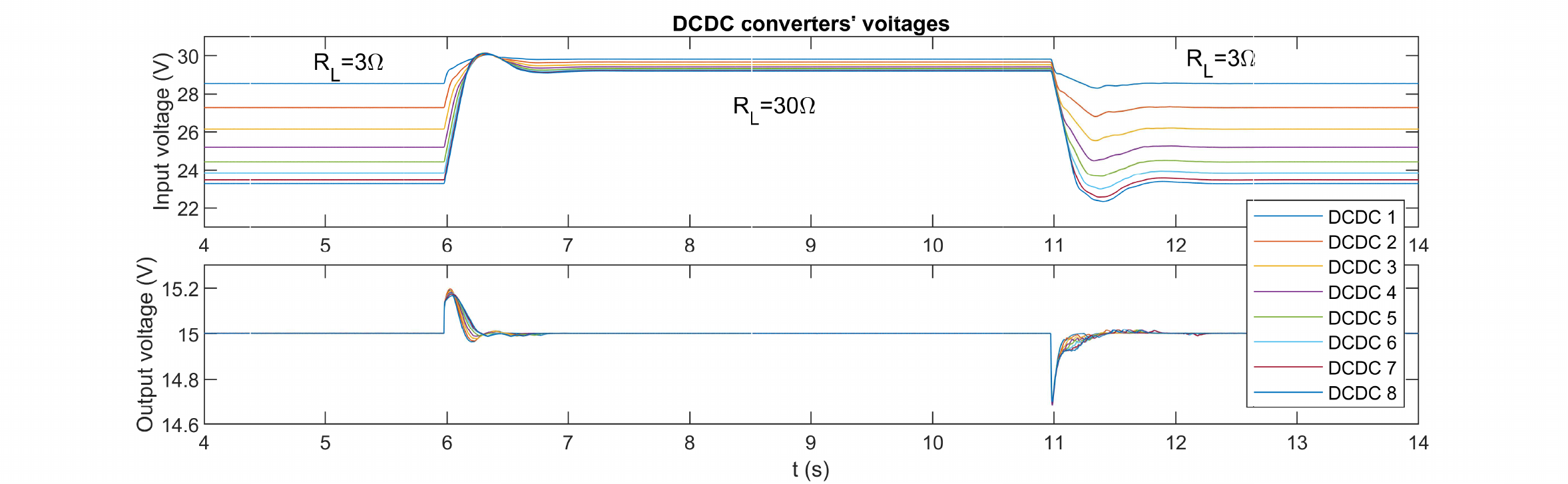}
   {(\textbf{a})}
    \includegraphics[bb=60bp 0bp 625bp 209bp,clip,width=1\linewidth]{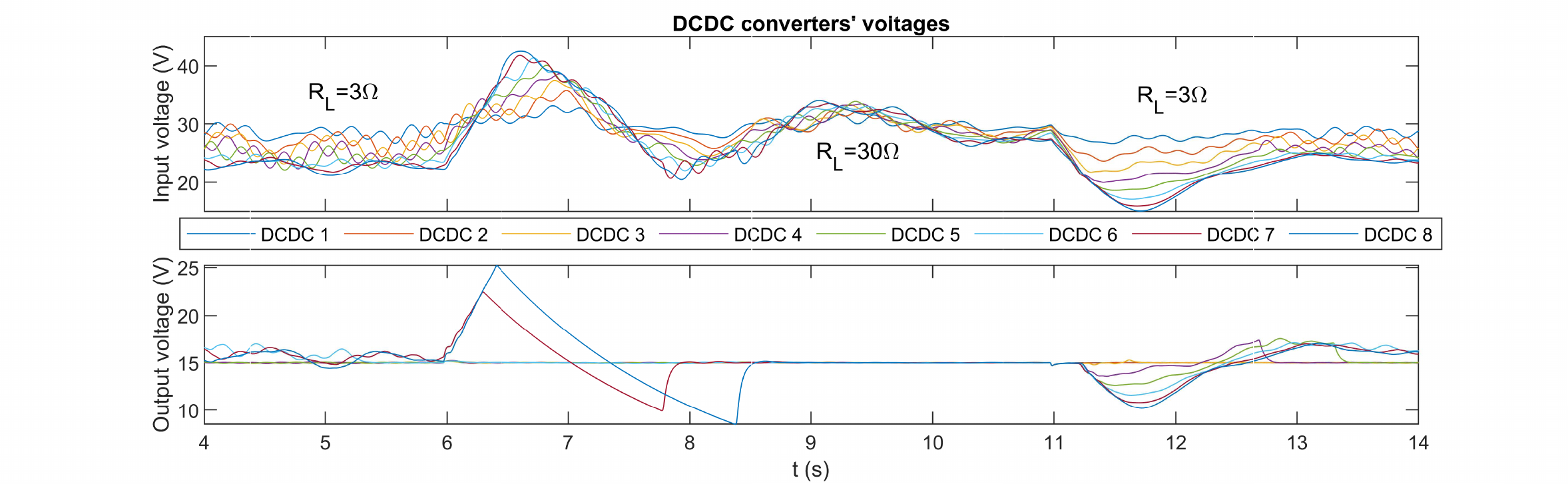}
    {(\textbf{b})}
    \caption{{Real-time simulation results for the radial network with $8$ buck converters: (\textbf{a}) using the line time-constants of $\tau=100$ ms, and (\textbf{b}) using $\tau=1000$ ms.}}\label{fig:rad}
\end{figure}

\begin{figure}
    \centering
    \includegraphics[bb=20bp 0bp 536bp 209bp,clip,width=1\linewidth]{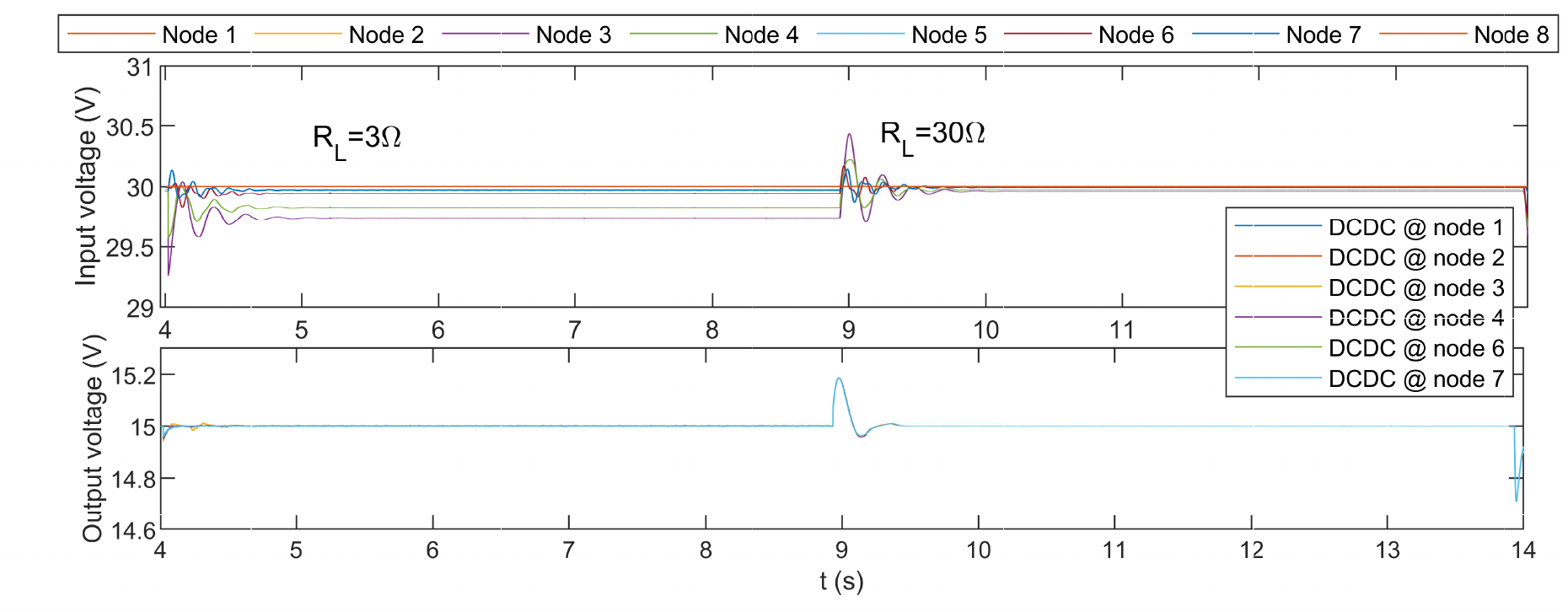}
   {(\textbf{a})}
    \includegraphics[bb=9bp 0bp 536bp 209bp,clip,width=1\linewidth]{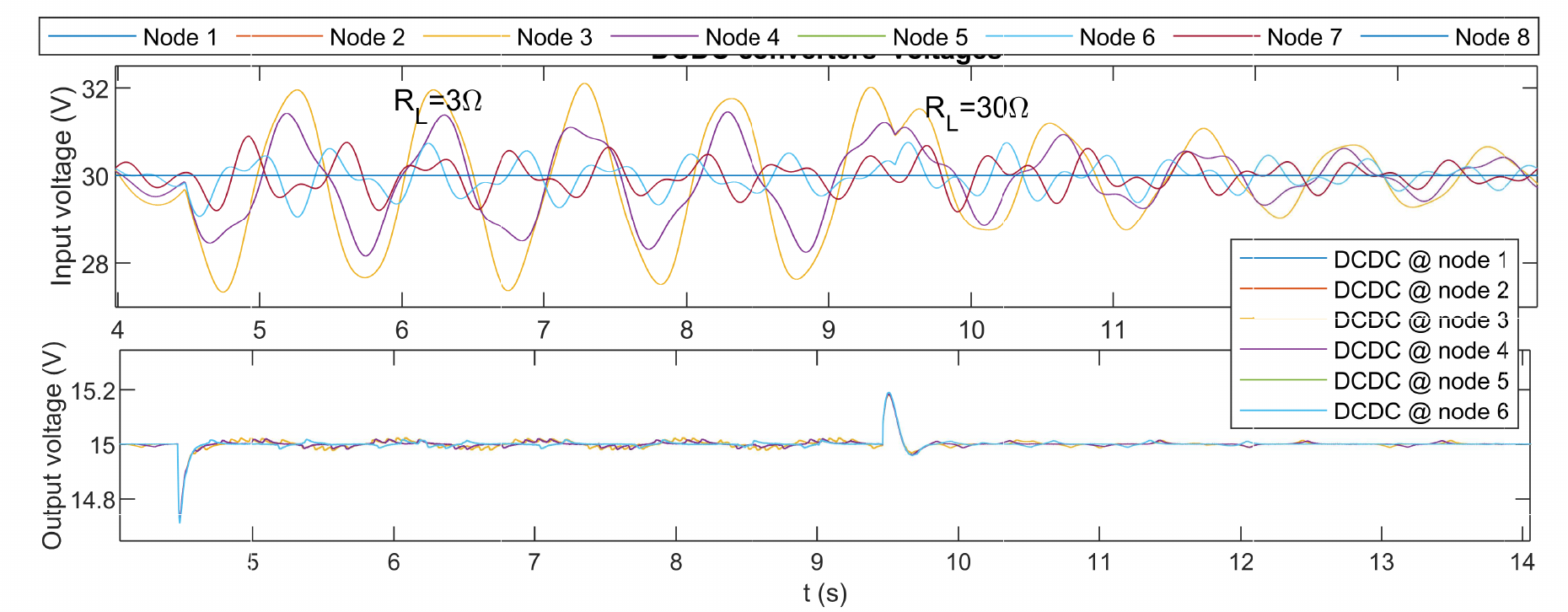}
    {(\textbf{b})}
    \caption{{Real-time simulation results for the meshed network of Fig. \ref{fig:sim_setup}:  using the line time-constants of $\tau=100$ ms, and (\textbf{b}) using $\tau=2000$ ms.}} \label{fig:mesh}
\end{figure}

{After performing the validation tests, we proceeded to test the stability of both radial and meshed networks with thew real-time simulations at $1$kHZ switching frequency. Fig. \ref{fig:rad} shows results for the radial grid with $8$ buck converters; specifically, input and output voltages of each converter are shown. Panel (\textbf{a}) show the simulation results for the line time-constant of $k_f\times1$ ms = $100$ ms, clearly indicating the stable operation. Panel (\textbf{b}) corresponds to the case of the time-constant $k_f\times10$ ms = $1000$ms. Under these conditions, the system already becomes unstable (or nearly unstable). We recall that our method already stops guaranteeing stability at the time-constant $~k_f\times6$ ms = $600$ms. Fig. \ref{fig:mesh} shows the real-time simulation results for the meshed topology of Fig. \ref{fig:sim_setup}. From the panel (\textbf{a}), we see, that for the case of line time-constant of $k_f\times1$ ms = $100$ ms the systems is stable, as in the previous radial case. Panel (\textbf{b}) shows the results for $k_f\times20$ ms = $2000$ ms, which indicates unstable system operation. In the case of the meshed network, the instability onset happens later than for the radial case; when the lines' time-constant is $k_f\times10$ ms = $1000$ ms, the system is still stable. We recall that our method indicates unstable operation might happen already at the line time-constants of $~k_f\times6$ ms = $600$ ms for some system configuration.}

\section{Practical Applications to the Design and Operation of DC Microgrids}\label{sec: design_operation}
Although mathematical in nature, the results of Corollary \ref{corollary: phase_simp2} can lead to a plethora of {1)} practical design guidelines for engineers and {2)} useful operational tools for non-expert microgrid users. Both are considered in the present section. 

\subsubsection{Microgrid Design Tools} If an engineering team desires network topology invariant compatibility of microgrid components, they can follow the steps of Algorithm \ref{alg:MDP}. {In this algorithm, a series of potential microgrid devices $\Sigma_1$, $\Sigma_2$, ... $\Sigma_i$ (e.g., lines, loads, and sources) are collectively\footnote{{The term ``collectively" indicates that the individual device dynamics are analyzed concurrently; it does not mean that a centralized eigenmode solver (e.g., (\ref{eq: s_vec})) is used to certify stability for any particular network configuration.}} analyzed across potential input voltages ${\rm V}_j$ and controller gains. If the phase responses of these devices pass the test stated on line 8, then any number of these devices can be interconnected in an arbitrary manner with a guarantee that the network will remain small-signal stable.} This procedure can have three outcomes: it will either (i) confirm the plug-and-play operability of all desired elements, (ii) determine the maximum vale of $\alpha$ for which the system can operate, {thus suggesting the controllers must be changed for guaranteed stable operation}, or (iii) identify the components which must be removed from the set in order for the system to operate at $\alpha=1$ (i.e., nominal controller values) {under arbitrary interconnection}.

Since the amount of overlap between phase curves implicitly parameterizes a stability margin, this procedure can also directly inform design constraints. {For instance, if the outcome of Algorithm $1$ shows stability for a certain set of devices, one can still scale the homotopy parameter $\alpha$ beyond unity to see at which value of $\alpha$ Algorithm $1$ fails. This will provide a certain indication of the robustness of control gains of the devices. For example, for buck converters with lead-lag voltage compensators with parameters from Table \ref{tab:sim_param_CC} and for lines with electro-magnetic time of $\tau=1$ ms, one can scale all the converters compensator gains by a factor of $10$ and still remain small-signal stable. This will provide the allowed range of the controller gain; however, in practice, additional control goals should also be considered (e.g., sufficient rejection of perturbations at certain frequencies). }

\begin{algorithm}
\caption{{Microgrid Stability Certification Procedure}}\label{alg:MDP}

 {\small \textbf{Require:}
Candidate components $\Sigma_i$ and potential input voltages ${\rm V}_j$

\begin{algorithmic}[1]

\For{each microgird component $\Sigma_i$}

\For{each potential input voltage ${\rm V}_j$}

\For{a sufficient number of $\alpha\in[0,1]$ values}

\State Parameterize closed loop gains of $\Sigma_i$ with $\alpha$

\State Compute \textbf{or} measure $\Sigma_i$'s phase response $\theta_i(\alpha,\!{\rm V}_{\!j},\omega)$

\EndFor {\bf end}

\EndFor {\bf end}

\EndFor {\bf end}

\For{each ascending value of $\alpha$}

\State Plot $\theta_{i}(\alpha,{\rm V}_{\!j},\omega)\pm90^{\circ},\;\forall i,\forall j,\forall\omega>0$

\If{there exists continuous overlap between \textbf{all} phase plots}

\State \textbf{All} components are plug-and-play stable for this $\alpha$

\Else

\State \textbf{either} terminate at this $\alpha$ \textbf{or} remove non-overlapping item

\EndIf {\bf end}

\EndFor {\bf end}

\State \Return Maximum $\alpha$ and non-rejected $\Sigma_i$, ${\rm V}_j$ pairs

\end{algorithmic}}
\end{algorithm}

\subsubsection{Microgrid Operational Tools} We consider a situation in which a variety of common microgrid components are collectively analyzed by Algorithm \ref{alg:MDP}. Immediate analysis could be performed by design engineers or researchers, while future analysis could be performed directly by manufacturers. Any particular combination of microgrid elements which satisfy the criteria of Algorithm \ref{alg:MDP} can then be aggregated to form a particular ``class". For user simplicity, each class can be associated with a color (i.e., ``red class", ``blue class", etc.). Of course, any particular microgird element can simultaneously belong to multiple classes, but mixing elements of certain classes could lead to instability. Depending on the needs of a microgrid, the designers could pre-determine the classes of elements which can be plug-and-play added to the microgrid without compromising the stability of the network. Users of the microgrid, then, would only need to check if a load, for example, satisfies the class designation of the microgrid. If it does, it can be safely added (i.e., plugged in) to any arbitrary location in the network without compromising stability.

{We would like to specifically point out an important feature of Algorithm \ref{alg:MDP}: the knowledge of network topology is not required to run it, just the knowledge of equivalent admittances of all the components, including lines (admittances can be even measured; there is no need to have models per se). After that, if Algorithm \ref{alg:MDP} is passed, components can be interconnected and small-signal stability of \emph{any} interconnection is automatically guaranteed without the need to run a separate stability test for every configuration. This is a distinct advantage of our proposed method compared to most other available methods.}

{\section{Discussion}\label{sec:discus}
The method developed in this manuscript allows for small-signal stability assessment of DC microgrids under arbitrary network topologies. In order to generate stability certificates, one needs to run the stability test for a set of components (including network lines) once, and, if the test is passed (i.e., the procedure of Algorithm \ref{alg:MDP}), components can be designated as compatible, and their arbitrary interconnection will be small-signal stable. Such a \emph{topology invariant} property is a distinct advantage of the method, compared to most other methods present in literature, since they typically require explicit stability assessment for every potential configuration.} 

{The present manuscript is the first one to introduce this method, and its main purpose is to show applicability to networks with arbitrary/unknown topologies, thus significantly simplifying a microgrid design procedure. Since the method is new, there are many aspects that may require additional research and validation, both in terms of the mathematical background and practical aspects. Below, we summarize the main advantages/contributions of the method and the main limitations/shortcomings, or points that need further research, in order to provide a fair analysis for readers.} 

{The advantages of the method follow:
\begin{itemize}
    \item The method naturally provides small-signal stability certificates for a set of components under arbitrary interconnection, thus eliminating the need to run separate stability analysis for every configuration;
    \item It can be applied to any type of component with any type of control settings, as long as effective admittance/impedance can be calculated or measured;
    \item Controller or PWM delays can be incorporated into the model by supplementing admittance functions with the corresponding exponential factors;
    \item In case stability cannot be certified, the method provides a principled way to change (i.e., lower) controller gains in order to guarantee system stability.
\end{itemize}}

{Following are the possible limitations/shortcomings or the points that need further research:
\begin{itemize}
    \item The method can certify small-signal stability only; additional research is needed to possibly incorporate some large signal transient stability aspects;
    \item For some networks, the method can be rather conservative, since by its virtue, it certifies stability of arbitrary interconnections of the given components; thus, it can be restrictive for some configurations;
    \item Some efficient mathematical methods should be developed for efficient assessment of the phase overlap condition \eqref{eq: phase_cond2} for the full range of frequencies and scaling factors $\alpha$;
    \item Additional research is needed to determine stability margins and compute robustness assessments in the method;
    \item Experimental validation of the method will greatly improve its validity for practical applications.
\end{itemize}}

{Since the method is new, additional advantages/challenges might be revealed by further research. However, even at the present state, the method demonstrates good potential and distinct advantages, many of which were confirmed by direct comparison with conventional stability assessment tools.}

\section{Conclusion}\label{sec:con}
{DC microgrids are a powerful and practical option for electrification of remote areas that the main grid cannot reach due to logistical or financial limitations.} In this paper, we have developed a fully decentralized small-signal stability criteria for DC networks that can allow for their plug-and-play operability. We have demonstrated that if a set of network elements meet certain criteria, then the system is guaranteed to be stable under arbitrary interconnection of the considered devices. We believe that these results can foster the development of ready-to-use standards for DC microgrids, and furthermore, that these standards can potentially be directly used by industry.

{While the theoretical underpinnings of the proposed method are rigorously sound, more research is needed in order to address both the practical implementation details and the methodology limitations. Accordingly, future work will begin testing the proposed plug-and-play stability standards in both hardware-in-the-loop (HIL) simulators and in physical low-voltage microgrid test-beds. In particular, we seek to understand the degree of conservativeness introduced by condition (\ref{eq: phase_cond2}) and whether or not this conservativeness can help safeguard the system from destabilizing large-signal disturbances.}

\appendices

{

\section{}\label{App_Init}
In order to assess small-signal stability, it is important to first compute the equilibrium point of the microgrid. To initialize the network, we assume steady state ($\omega=0$) operation. We then order the network such that buses $1$ through $n_{s}$ are source buses while buses $n_{s}+1$ through $n$ are load buses. We define $Y_l$ as the diagonal matrix of line admittances, where each diagonal element is $1/r_{ij}$, and we define $Y_s$ as the diagonal matrix of shunt (i.e., load) admittances, where each diagonal element is $D^2/R$ (i.e., the \textit{negative} of incremental resistance (\ref{eq: YL_SS})). We further define
\begin{align}\label{eq: Yb}
Y_{b}=E^{\top}Y_{l}E+Y_{s}
\end{align}
as the nodal admittance matrix. Next, we define ${\bm V}_s$ as the vector of \textit{known} source voltages, ${\bm V}_l$ as the vector of \textit{unknown} load voltages, and ${\bm I}_s$ as the vector of \textit{unknown} current injections at the sources. Partitioning (\ref{eq: Yb}), we therefore have
\begin{align}\label{eq: init_DC}
\left[\begin{array}{c|c}
Y_{b1} & Y_{b2}\\
\hline Y_{b3} & Y_{b4}
\end{array}\right]\left[\begin{array}{c}
\bm{V}_{s}\\
\hline \bm{V}_{l}
\end{array}\right]=\left[\begin{array}{c}
\bm{I}_{s}\\
\hline {\bf 0}
\end{array}\right].
\end{align}
By re-arranging the bottom set of equations from (\ref{eq: init_DC}), the unknown steady state voltages may be computed as
\begin{align}\label{eq: load_voltage}
\bm{V}_{l}=-Y_{b4}^{-1}Y_{b3}\bm{V}_{s}.
\end{align}

To introduce variety into the radial and meshed networks that we test, the line lengths are chosen from a normal distribution with a small standard deviation: $l_{ij}=1+{\mathcal N}(0,0.1^2)$. We then solve (\ref{eq: load_voltage}) using the parameter values provided in Table \ref{tab:sim_param_CC}.

\section{}\label{Boost_Converter}
{
We consider a boost converter that operates in the average current control mode (ACM) with a voltage compensation loop. For this mode, the following four transfer-function are important (see. \cite{erickson2007fundamentals}, Table $12.4$):
\begin{equation}\label{Gvd}
    G_{vd}=\frac{V}{D'} \frac{(1-s\frac{L}{D'^2 R})}{1+\frac{s}{Q\omega_0} + \left( \frac{s}{\omega_0}\right)^2}
\end{equation}
\begin{equation}\label{Gid}
    G_{id}=\frac{2V}{D'^2 R} \frac{(1+s\frac{RC}{2})}{1+\frac{s}{Q\omega_0} + \left( \frac{s}{\omega_0}\right)^2}
\end{equation}
\begin{equation}\label{Gvg}
    G_{vg}=\frac{1}{D'} \frac{1}{1+\frac{s}{Q\omega_0} + \left( \frac{s}{\omega_0}\right)^2}
\end{equation}
\begin{equation}\label{Gig}
    G_{ig}=\frac{1}{D'^2 R} \frac{(1+sRC)}{1+\frac{s}{Q\omega_0} + \left( \frac{s}{\omega_0}\right)^2}.
\end{equation}
Current compensator $G_{ci}$ and voltage compensator $G_{cv}$ are given by the following relations:
\begin{equation}\label{Gci}
    G_{ci}=G_{cm}\frac{(1+\frac{\omega_{c1}}{s})}{(1+\frac{s}{\omega_{c2}})}
\end{equation}
\begin{equation}\label{Gcv}
    G_{cv}=G_{vm}\left(1+\frac{\omega_{v1}}{s} \right).
\end{equation}
The functions written above can be used to calculate the boost converter small-signal input admittance:
\begin{equation}\label{Y_boost}
    Y_{\rm boost}=\frac{G_{ig}+ G_{cv}G_{ci}G_{vd}G_{ig}- G_{cv}G_{ci}G_{id}G_{vg} }{1+ G_{ci} G_{id} + G_{cv} G_{ci} G_{vd}} ,
\end{equation}
where the expressions for transfer functions $G$ in \eqref{Y_boost} are given by \eqref{Gvd}-\eqref{Gcv}. In our experiments, the boost converter was connected in parallel with a grounded filter capacitor with $C_f=2500$ $\mu$F. The other parameters associated with the boost converter model are given by $R=33.33$ $\Omega$ , $C=500$ $\mu$F, $L=50$ $\mu$H, $D'=0.56$, $G_{cm}=0.0318$, $G_{vm}=3.125$, $\omega_{c1}=2\pi\cdot400$ Hz, $\omega_{c2}=2\pi\cdot12.5$ kHz, and $\omega_{c2}=2\pi\cdot167$ Hz. The nominal input voltage was $V=28$ V.}

\bibliographystyle{ieeetr}
\bibliography{mainbib}

\end{document}